\documentclass[11pt]{article}
\pdfoutput=1
\usepackage[centertags]{amsmath}
\usepackage[square, comma, sort&compress,numbers]{natbib}
\usepackage{array,multirow}
\numberwithin{equation}{section}
\usepackage{amssymb,amsfonts}
\usepackage{graphicx}
\usepackage{color}
\usepackage{mathtools,bm}
\usepackage{accents}

\usepackage{epsfig}
\usepackage{bbold}

\usepackage{wrapfig}
\usepackage{float}
\usepackage{soul}
\usepackage{tkz-euclide}

\usepackage{braket}

\usepackage{tikz,pgf}
\usetikzlibrary{shapes}
\usetikzlibrary{calc}
\usetikzlibrary{decorations.pathmorphing}
\usetikzlibrary{decorations.pathreplacing,shapes.misc}
\usetikzlibrary{positioning}
\usetikzlibrary{arrows}
\usetikzlibrary{decorations.markings}
\usetikzlibrary{shadings}

\usetikzlibrary{intersections}

\def\be{\begin{equation}}
\def\ee{\end{equation}}
\def\ba{\begin{array}}
\def\ea{\end{array}}

\def\dps{\displaystyle}

\def\1{\tilde{1}}
\def\2{\tilde{2}}
\def\3{\tilde{3}}

\newdimen\tableauside\tableauside=1.0ex
\newdimen\tableaurule\tableaurule=0.4pt
\newdimen\tableaustep
\def\phantomhrule#1{\hbox{\vbox to0pt{\hrule height\tableaurule
width#1\vss}}}
\def\phantomvrule#1{\vbox{\hbox to0pt{\vrule width\tableaurule
height#1\hss}}}
\def\sqr{\vbox{%
\phantomhrule\tableaustep

\hbox{\phantomvrule\tableaustep\kern\tableaustep\phantomvrule\tableaustep}%
\hbox{\vbox{\phantomhrule\tableauside}\kern-\tableaurule}}}
\def\squares#1{\hbox{\count0=#1\noindent\loop\sqr
\advance\count0 by-1 \ifnum\count0>0\repeat}}
\def\tableau#1{\vcenter{\offinterlineskip
\tableaustep=\tableauside\advance\tableaustep by-\tableaurule
\kern\normallineskip\hbox
{\kern\normallineskip\vbox
{\gettableau#1 0 }%
\kern\normallineskip\kern\tableaurule}%
\kern\normallineskip\kern\tableaurule}}
\def\gettableau#1 {\ifnum#1=0\let\next=\null\else
\squares{#1}\let\next=\gettableau\fi\next}

\newcommand{\bref}[1]{\textbf{\ref{#1}}}

\newcommand{\im}{\mathop{\mathrm{Im}}}
\newcommand{\re}{\mathop{\mathrm{Re}}}

\def\cF{\mathcal{F}}

\def\cH{\mathcal{H}}

\def\cL{\mathcal{L}}

\def\cO{\mathcal{O}}

\def\cV{\mathcal{V}}

\numberwithin{equation}{section} \makeatletter
\@addtoreset{equation}{section}

\def\be{\begin{equation}}
\def\ee{\end{equation}}
\def\ba{\begin{array}}
\def\ea{\end{array}}

\def\dps{\displaystyle}

\def\ba{\begin{array}}
\def\ea{\end{array}}

\def\dps{\displaystyle}

\def \bz{\bar z}

\def\cft{CFT$_d$ }

\def\bz{{\bf z}}

\def\wb{\bar{{w}}}
\def\zb{\bar{z}}

\newcommand{\dl}{\Delta}

\newcommand{\dlb}{\bar{\Delta}}

\newcommand{\hb}{\bar{h}}

\def\C2{\text{C}_2}

\def\cfttwo{CFT$_2$ } 
\def\cftone{CFT$_1$ }

\newcommand{\kostya}[1]{{\color{purple}#1}}

\def\?{\kostya{???}\;}

\usepackage{jheppub}
\makeatletter
\def\@fpheader{\vspace{-.1cm}}
\makeatother

\title{\centering{Torus shadow formalism and \\ exact global conformal blocks}}

\author{Konstantin\ Alkalaev and}
\author{Semyon\ Mandrygin}

\affiliation{I.E. Tamm Department of Theoretical Physics, \\P.N. Lebedev Physical
Institute, 119991 Moscow, Russia}

\emailAdd{alkalaev@lpi.ru}
\emailAdd{semyon.mandrygin@gmail.com}

\abstract{Using the shadow formalism  we find  global conformal blocks of torus CFT$_2$. It is shown that  $n$-point torus blocks  in the ``necklace'' channel (a loop with $n$ legs) are expressed in terms of a  hypergeometric-type  function which we refer to as the necklace function. 
}

\begin{document}

\maketitle
\flushbottom

\section{Introduction}

The shadow formalism in conformal field theory \cite{Ferrara:1972uq,Ferrara:1972kab} has proved to be an efficient tool to calculate conformal blocks, see e.g. \cite{SimmonsDuffin:2012uy,Rosenhaus:2018zqn}. In this paper we introduce the  shadow formalism for torus \cfttwo and compute global $n$-point  conformal blocks in the necklace channel.\footnote{Some partial results were previously obtained in \cite{Hadasz:2009db, Alkalaev:2016fok, Kraus:2017ezw, Alkalaev:2017bzx, Cho:2017oxl, Gerbershagen:2021yma, Alkalaev:2022kal, Pavlov:2023asi}.  Torus blocks in BMS \cfttwo are discussed in \cite{Bagchi:2020rwb}, the shadow formalism in Galilean \cfttwo is considered in \cite{Chen:2022cpx}. Also, the large-$c$ torus blocks can be computed through the AdS$_3$/CFT$_2$ correspondence as the lengths of particular graphs stretched in the bulk having the rigid  torus topology \cite{Alkalaev:2016ptm,Kraus:2017ezw,Alkalaev:2017bzx,Gobeil:2018fzy,Brehm:2018ipf,RamosCabezas:2020mew,Alkalaev:2020yvq}.}      

\begin{figure}
\center{\includegraphics[width=0.4\linewidth]{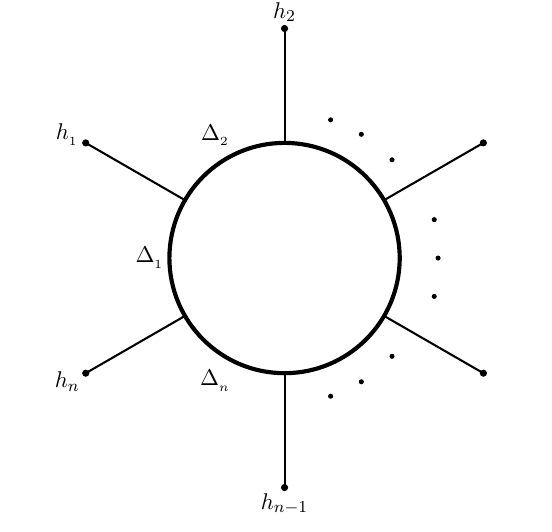}}
\caption{The $n$-point torus block in the necklace channel. The primary operators are in $\{z_1,..., z_n\}$; the external dimensions are $\{h_1, ... , h_n\}$ and the intermediate dimensions are $\{\Delta_1, ... , \Delta_{n}\}$.}
\label{fig:neck}
\end{figure}

Let us formulate our main result. In torus \cfttwo  the global conformal block in the necklace channel (see Fig. \bref{fig:neck}) reads 
\be
\cF_{{\bf \dl}}^{{\bf h}}(q,\bz)\;\sim\; 
F_N \left[
\begin{array}{l l}
a_1, ..., a_n \\
c_1, ..., c_n
\end{array}\bigg| 
\rho_1,..., \rho_n 
\right],\qquad n=2,3,...\,,
\ee
where  
\be
\label{neckalcefunction0}
F_N \left[
\begin{array}{l l}
a_1, ..., a_n \\
c_1, ..., c_n
\end{array}\bigg| 
\rho_1,..., \rho_n 
\right]
= \sum_{m_1,... ,m_n = 0}^{\infty} \frac{(a_1)_{m_1+m_2} (a_2)_{m_2+m_3} ... (a_n)_{m_n+m_1} }{(c_1)_{m_1} ... (c_n)_{m_n} } \frac{\rho_1^{m_1} }{m_1!} ... \frac{\rho_n^{m_n} }{m_n!} \,.
\ee
The parameters $a_i$ and $c_i$ are linear functions of conformal dimensions; the variables $\rho_i$ are particular combinations of $z$-coordinates and the modular parameter $q$; the symbol $\sim$ means that there is a prefactor called the leg factor which defines the conformal transformation properties of the torus block; the function $F_N$ is the necklace function which defines the bare conformal block. The case of 1-point blocks is exceptional.

The  paper is organized as follows. In Section \bref{sec:global} we describe  the shadow formalism: in Section \bref{ShadowIn2Dim} we review the shadow formalism in  plane  CFT$_2$; in Section \bref{sec:4pointBlock}, following \cite{Rosenhaus:2018zqn} we demonstrate how to apply this formalism to calculate the 4-point conformal block in the comb channel focusing on all basic features needed further in torus CFT$_2$; the known results for multipoint conformal blocks in the comb channel expressed in terms of the comb function are summarized in Section \bref{sec:n_p}. In Section \bref{sec:TorusBl} we formulate the shadow formalism for torus CFT: in Section \bref{sec:BasicsTorus} we  set our notation and conventions for torus \cfttwo and review the $n$-point torus correlation functions; in Section \bref{sec:toruscpw} we define torus \cfttwo conformal partial waves. Here, using the (anti)chiral factorization we reduce the whole analysis to the one-dimensional case where the torus is replaced by a circle. The main is Section \bref{sec:calc}, where we calculate multipoint torus conformal blocks: in Section \bref{1pointTorusBlock} we calculate both 0-point block (character) and 1-point block; Section \bref{sec:torus2} explicitly describes our basic example of the 2-point torus block which is shown to be defined by the forth Appell function; in Section \bref{sec:torusmulti} we extend this calculation scheme to the case of $n$ points and introduce the necklace function.     In Section \bref{sec:concl} we discuss our results and possible further developments.  Appendix \bref{app:multiblocks} collects the main formulae for $n$-point conformal blocks in the comb channel of plane \cfttwo and reviews the comb function \cite{Rosenhaus:2018zqn}.  Appendix \bref{app:necklacefunction} introduces the necklace function and describes its properties.

\section{Global plane conformal blocks }
\label{sec:global}

\subsection{Shadow formalism}
\label{ShadowIn2Dim}

For any  primary operator $\cO(z, \zb)$ with (anti-)holomorphic dimensions $\dl,\dlb$ one can  introduce  a {\it shadow  operator} which is a primary operator with (anti)holomorphic dimensions $1-\dl$, $1-\dlb$ \cite{Ferrara:1972kab,Ferrara:1972uq}:
\be
\label{ShadowOp}
\widetilde{\cO}(z,\zb) = \frac{1}{N_{\dl,\dlb}}\int_{\mathbb{R}^2} {\rm d}^2 w\, \frac{\cO(w,\wb)}{(z-w)^{2(1-\dl)} (\zb-\wb)^{2(1-\dlb)}}  \,,
\ee
where the measure  is ${\rm d}^2 w \equiv  {\rm d} \left( \re w \right) \, {\rm d} \left( \im w \right)$. The shadow operator \eqref{ShadowOp} realizes an equivalent representation for $\cO$ that means that \cite{Fradkin:1996is, Ferrara:1972uq}:\footnote{\label{fn1}The $\delta$-function is defined by  $\int_{\mathbb{R}^2} {\rm d}^2 w f(w,\wb) \delta^2 (z-w) = f(z,\zb) $.}
\be
\label{shadow2pt}
\big\langle \widetilde{\cO}(z,\zb) \cO(w,\wb)\big\rangle = \delta^2 (z-w) \,,
\ee
Equation \eqref{shadow2pt} fixes the normalization $N_{\dl,\dlb}$ in \eqref{ShadowOp} (see \cite{Dolan:2011dv} for evaluating the integral):
\be
\label{NormShadow}
N_{\dl,\dlb}=\pi^2(-)^{2(\dl-\dlb)}\frac{\Gamma(2\dl-1) \Gamma(1-2\dl) }{\Gamma(2-2\dlb)) \Gamma(2\dlb) } \,,
\ee
however, there are other normalization conventions, see \cite{Osborn:2012vt, Fradkin:1996is, Anous:2020vtw}. We remind that $\dl-\dlb=s_{\cO}\in \mathbb{Z}/2$, $\dl+\dlb=\dl_{\cO}$ is a spin and a conformal weight of $\cO$, respectively. 

Using shadow operators one can define projectors 
\be
\label{projector}
\Pi_{\dl,\dlb} =  \int _{\mathbb{R}^2} {\rm d}^2 z\,  \cO(z,\zb) \ket{0} \bra{0} \widetilde{\cO}(z,\zb) \,,
\ee
which obey the standard idempotent  and orthogonality relations (which directly follow from \eqref{shadow2pt})
\be
\label{projprop}
\Pi_{\dl_m,\dlb_m} \Pi_{\dl_n,\dlb_n} = \delta_{mn}\Pi_{\dl_m,\dlb_m} \,.
\ee
The projectors are conformally invariant, i.e. the  integrand in \eqref{projector} has the  (anti)holomorphic dimension $1$ under conformal transformations of $z,\zb$. 

Let us  consider  the 4-point correlation function of primary operators $\phi_{h_i,\bar{h}_i}(z_i, \bar z_i) \equiv \phi_i$. Inserting the projector \eqref{projector}  between the pairs of operators $\phi_1 \phi_2$ and $\phi_3 \phi_4$ one defines the conformal partial wave (CPW)
\be
\ba{c}
\label{cpw4full}
\dps
\Psi_{\dl,\dlb}^{h_1,...,h_4 ; \bar{h}_1,..., \bar{h}_4}(z_1,\zb_1,...\,, z_4,\zb_4) = \int_{\mathbb{R}^2} {\rm d}^2 w \,V_{h_1,h_2,\dl}(z_1,z_2,w) V_{1-\dl,h_3,h_4}(w,z_3,z_4)  
\vspace{2mm} 
\\ 
\hspace{71mm}\times\, \bar{V}_{\bar{h}_1,\bar{h}_2,\dlb}(\zb_1,\zb_2,\wb) \bar{V}_{1-\dlb,\bar{h}_3,\bar{h}_4}(\wb,\zb_3,\zb_4) \,,
\ea
\ee
where we introduced the notation 
\be
\label{vertex}
V_{h_i,h_j,h_k}(z_i,z_j,z_k) = \frac{1}{(z_{ij})^{h_i+h_j-h_k} (z_{ik})^{h_i+h_k-h_j} (z_{jk})^{h_j+h_k-h_i}} \,,
\ee
where $z_{ij}\equiv z_{i,j} = z_i-z_j$. Then, the 3-point function of primary operators $\phi_i$ reads as 
\be
\big\langle \phi_i (z_i,\zb_i) \phi_j (z_j,\zb_j) \phi_k (z_k, \zb_k) \big\rangle = C_{ijk}\, V_{h_i,h_j,h_k} (z_i,z_j,z_k) \bar{V}_{\bar{h}_i,\bar{h}_j,\bar{h}_k} (\zb_i,\zb_j,\zb_k) \,,
\ee
where $C_{ijk}$ are the structure constants,  and $\bar{V}$ is obtained form $V$ by substituting $z_{i,j,k} \rightarrow \zb_{i,j,k}$ and $h_{i,j,k} \rightarrow \bar{h}_{i,j,k}$. The CPW \eqref{cpw4full} is given by a sum of the conformal block and its shadow \cite{Osborn:2012vt,SimmonsDuffin:2012uy} 
\be
\begin{split}
\label{cpw4fullsum}
\dps
\Psi_{\dl,\dlb}^{h_1,... , h_4 ; \bar{h}_1,..., \bar{h}_4}(z_1,\zb_1 , ...\, , z_4, \zb_4) 
&= \alpha \, G_{\dl,\dlb}^{h_1,..., h_4 ; \bar{h}_1, ..., \bar{h}_4 } (z_1, \bar{z}_1,...\, , z_4, \zb_4 )
\\ 
&+ \beta \, G_{1-\dl,1-\dlb}^{h_1,..., h_4 ; \bar{h}_1, ..., \bar{h}_4 } (z_1,\zb_1 , ...\, , z_4, \zb_4) \,,
\end{split}
\ee
with some coefficients $\alpha, \beta$ depending on external/intermediate conformal dimensions $h_i,\dl$.  In \cfttwo the conformal block factorizes into (anti)holomorphic parts
\be
\label{blockfactorization}
G_{\dl,\dlb}^{h_1,..., h_4 ; \bar{h}_1, ..., \bar{h}_4 } (z_1, \bar{z}_1, ...\, , z_4, \zb_4) = G_{\dl}^{h_1,h_2,h_3,h_4}(z_1, z_2, z_3, z_4)\, \bar{G}_{\dlb}^{\bar{h}_1, \bar{h}_2, \bar{h}_3, \bar{h}_4 } (\bar{z}_1, \bar{z}_2, \bar{z}_3, \zb_4) \,.
\ee
The same formula is valid for the shadow block $G_{1-\dl,1-\dlb}^{h_1,\ldots h_4 ; \bar{h}_1, \ldots, \bar{h}_4 }$. The holomorphic shadow conformal block  $G_{1-\dl}^{h_1,\ldots ,h_4}$ is obtained from the holomorphic conformal block by  $\dl \rightarrow 1-\dl$.

When all points $z_i \in \mathbb{R}$ the holomorphic block  can be seen as \cftone conformal block. Thus, the full \cfttwo CPW \eqref{cpw4fullsum} can be represented by knowing only one-dimensional  conformal blocks. Having this in mind, one can define the one-dimensional  CPW \cite{Rosenhaus:2018zqn}:
\be
\label{cpw4}
\Psi_{\dl}^{h_1,h_2,h_3,h_4}(z_1,z_2,z_3,z_4) = \int_{\mathbb{R}} {\rm d}w\, V_{h_1,h_2,\dl}(z_1,z_2,w) V_{1-\dl,h_3,h_4}(w,z_3,z_4) \,,
\ee
where from now on $z_i \in \mathbb{R}$. This function is a sum of the conformal block and its shadow,
\be
\label{cpw4sum}
\Psi_{\dl}^{h_1,h_2,h_3,h_4}(z_1,z_2,z_3,z_4) = a\, G_{\dl}^{h_1,h_2,h_3,h_4}(z_1, z_2, z_3 , z_4)+ b\,
G_{1-\dl}^{h_1,h_2,h_3,h_4}(z_1, z_2, z_3 , z_4 )\,,
\ee
with  some coefficients $a ,b$ depending on external/intermediate conformal dimensions.
In order to distinguish between conformal and shadow  blocks one notes that the conformal block $G_{\dl}$ has a particular asymptotic behaviour in the OPE limit \cite{Rosenhaus:2018zqn}\footnote{In particular, this asymptotics follows from the known exact expression for the 4-point conformal block
\be
\label{block4old}
G_{\dl}^{h_1,h_2,h_3,h_4}(z_1, z_2, z_3 , z_4) =\frac{1}{z_{12}^{h_1+h_2} z_{34}^{h_3+h_4}} \left(\frac{z_{24}}{z_{14}} \right)^{h_1-h_2} \left(\frac{z_{14}}{z_{13}} \right)^{h_3-h_4} \chi_1^{\dl}\; {}_2 F_1  
\left[
\begin{array}{l l}
\dl-h_{12}, \dl+h_{34} \\
\qquad 2\dl
\end{array}\bigg| \; \chi_1
\right] \,,
\ee
where  $\chi_1 = (z_{12} z_{34})/(z_{13} z_{24})$ is the  cross-ratio  \cite{Belavin:1984vu}.}  
\be
\label{boundary4}
G_\dl ^{h_1,h_2,h_3,h_4} \rightarrow z_{34}^{\dl-h_3-h_4} V_{h_1,h_2,\dl}(z_1,z_2,z_3) \quad \text{at} \quad z_3 \to z_4 \,.
\ee
Finally,  substituting the 3-point functions \eqref{vertex} into the   CPW \eqref{cpw4} one finds the following integral representation for the one-dimensional  CPW:\footnote{Note that when considering  \eqref{vertex} in $d=1$  one substitutes $z_{ij}\rightarrow |z_{ij}|$.}
\be
\label{cpw4final}
\Psi_{\dl}^{h_1,h_2,h_3,h_4}(z_1,z_2,z_3,z_4) = \int_{\mathbb{R}} {\rm d}w\, \frac{|z_{12}|^{-h_1-h_2+\dl} |z_{34}|^{-h_3-h_4+1-\dl}}{|w-z_1|^{\dl+h_{12}} |w-z_2|^{\dl-h_{12}} |w-z_3|^{1-\dl+h_{34}} |w-z_4|^{1-\dl-h_{34}} }\,,
\ee
where $h_{kl} \equiv   h_k-h_l$.

\subsection{4-point conformal block}
\label{sec:4pointBlock}

By making  a suitable change of the integration variable the integral \eqref{cpw4final} can be reduced to 
\begin{multline}
\label{cpwcross4}
\Psi_{\dl}^{h_1,h_2,h_3,h_4}(z_1,z_2,z_3,z_4) = \frac{1}{|z_{12}|^{h_1+h_2} |z_{34}|^{h_3+h_4} } \left| \frac{z_{24}}{z_{14}} \right|^{h_{12}}
\left| \frac{z_{14}}{z_{13}} \right|^{h_{34}} 
\\
\times \int_{\mathbb{R}} {\rm d}w\, \frac{\chi_1^{1-\dl}}{|w|^{1-\dl-h_{34}} |1-w|^{\dl-h_{12}} |\chi_1-w|^{1-\dl+h_{34}} } \,.
\end{multline}
Here, the cross-ratio $\chi_1 = (z_{12} z_{34})/(z_{13} z_{24}) < 1$ that corresponds to ordering points as $z_1>z_2>z_3>z_4$. The integral in \eqref{cpwcross4} has five singular points $w=\{-\infty,0,\chi_1,1,+\infty \}$ and can be evaluated by splitting the integration domain into four regions as  
\be
\label{4pointSplit}
\int_{\mathbb{R}} {\rm d}w \, I(\chi_1,w)  
=  \int_{-\infty}^{0} {\rm d}w\, I(\chi_1,w)   + \int_{0}^{\chi_1} {\rm d}w\, I(\chi_1,w)  
+ \int_{\chi_1}^{1} {\rm d}w\, I(\chi_1,w)  +  \int_{1}^{+\infty} {\rm d}w\, I(\chi_1,w) \,,
\ee
where the integrand is given by 
\be
I(\chi_1,w) = \frac{\chi_1^{1-\dl}}{|w|^{1-\dl-h_{34}} |1-w|^{\dl-h_{12}} |\chi_1-w|^{1-\dl+h_{34}} } \,.
\ee
The integrals \eqref{4pointSplit} are found to be: 
\be
\label{cpw4integrals}
\begin{split}
\int_{-\infty}^{0} {\rm d}w\, I(\chi_1,w)  &= a_1\, \mathfrak{I}_{\dl}^{h_1,h_2,h_3,h_4}(\chi_1) +b_1\, \mathfrak{I}_{1-\dl}^{h_1,h_2,h_3,h_4}(\chi_1) \,,
\\
\int_{0}^{\chi_1} {\rm d}w\, I(\chi_1,w)  &=  a_2\, \mathfrak{I}_{\dl}^{h_1,h_2,h_3,h_4}(\chi_1) \,,
\\
\int_{\chi_1}^{1} {\rm d}w\, I(\chi_1,w) &= a_3\, \mathfrak{I}_{\dl}^{h_1,h_2,h_3,h_4}(\chi_1) + b_3\, \mathfrak{I}_{1-\dl}^{h_1,h_2,h_3,h_4}(\chi_1) \,,
\\
\int_{1}^{+\infty} {\rm d}w\, I(\chi_1,w)  &= b_4\, \mathfrak{I}_{1-\dl}^{h_1,h_2,h_3,h_4}(\chi_1) \,,
\end{split}
\ee
where the coefficients $a_i$, $b_i$ depend  on conformal dimensions $h_i,\dl$ only, and
\be
\mathfrak{I}_{\dl}^{h_1,h_2,h_3,h_4}(\chi_1) = \chi_1^{\dl}\,  {}_2 F_1  
\left[
\begin{array}{l l}
\dl-h_{12}, \dl+h_{34} \\
\qquad 2\dl
\end{array}\bigg| \; \chi_1
\right].
\ee
In any integration domain the integrals \eqref{cpw4integrals} involve 2 functions $\mathfrak{I}_{\dl}$ and $\mathfrak{I}_{1-\dl}$ related by   $\dl \rightarrow 1-\dl$, they have  different asymptotics at $\chi_1 \rightarrow 0$ (i.e. $z_3 \to z_4$). From \eqref{cpw4sum} we conclude that these two functions contribute either to the conformal block $G_{\dl}$ or to the shadow block $G_{1-\dl}$. Assuming that the correct  asymptotics is given by  \eqref{boundary4} one concludes that $\mathfrak{I}_{\dl}$ contributes to the conformal block $G_{\dl}$. In particular, knowing the asymptotics allows relating  the coefficients in   \eqref{cpw4sum} and \eqref{cpw4integrals} as  $a = \sum_i a_i$ and  $b=\sum_i b_i$. 

As noted in \cite{Rosenhaus:2018zqn}, in order to find $G_{\dl}$ there is no need to integrate over the whole $\mathbb{R}$ in \eqref{cpwcross4}. Instead, the above splitting of the integration domain and the formulas \eqref{cpw4integrals} allows one to single out the conformal block contribution by choosing a suitable integration interval. E.g. choosing the interval $(0, \chi_1)$ one directly finds the conformal block. On the other hand, choosing the interval $(1,+\infty)$ one finds the shadow block which can be converted to the conformal block by $\Delta \to 1-\Delta$. For other two intervals the result is given by a linear combination of the conformal and shadow blocks which can be disentangled by checking their asymptotics.  Whatever the integration interval is chosen the 4-point conformal block is calculated to be 
\be
\label{block4}
G_{\dl}^{h_1,h_2,h_3,h_4}(z_1,z_2,z_3,z_4) = \mathcal{L}^{h_1,h_2,h_3,h_4}(z_1,z_2,z_3,z_4)\, g_{\dl}^{h_1,h_2,h_3,h_4}(\chi_1)\,.
\ee
Here,  one  introduces the leg factor \cite{Rosenhaus:2018zqn}: 
\be
\label{leg4}
\mathcal{L}^{h_1,h_2,h_3,h_4}(z_1,z_2,z_3,z_4) = \frac{1}{|z_{12}|^{h_1+h_2} |z_{34}|^{h_3+h_4} } \left| \frac{z_{23}}{z_{13}} \right|^{h_{12}}
\left| \frac{z_{24}}{z_{23}} \right|^{h_{34}},
\ee
which guarantees correct conformal transformation properties of the conformal block; the function  $g_{\dl}^{h_1,h_2,h_3,h_4}(\chi_1)$ is  the bare 4-point conformal block which,  by definition, is a conformally invariant  function,
\be
\label{blockbare4}
g_{\dl}^{h_1,h_2,h_3,h_4}(\chi_1) = \chi_1^{\dl}\,  {}_2 F_1  
\left[
\begin{array}{l l}
\dl + h_{12}, \dl - h_{34} \\
\qquad 2\dl
\end{array}\bigg| \; \chi_1
\right].
\ee
Note that the bare conformal block in  this form differs from the common one \eqref{block4old}. Their relation can be seen by using  the Euler identity 
\be
{}_2 F_1 
\left[
\begin{array}{l l}
\dl-h_{12},\dl+h_{34} \\
\qquad  2\dl
\end{array}\bigg| \; \chi_1
\right]  = (1-\chi_1)^{h_{12}-h_{34}} {}_2 F_1 
\left[
\begin{array}{l l}
\dl+h_{12},\dl-h_{34} \\
\qquad  2\dl
\end{array}\bigg| \; \chi_1
\right].
\ee
Of course, these two representations are equivalent. However,  splitting the conformal block into the leg factor and bare conformal block is ambiguous. In general, they are related by a factor of a function of the cross-ratios.  The point of making the Euler transform is to choose a particular  leg factor that would allow to represent multipoint  bare blocks in the simplest form.  In this respect, the leg factor \eqref{leg4} is preferable  than the leg factor in \eqref{block4old}.

\subsection{Multipoint  conformal blocks}
\label{sec:n_p}

\begin{figure}
\center{\includegraphics[width=0.6\linewidth]{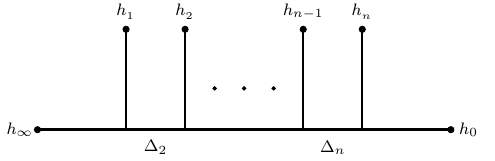}}
\caption{The $(n+2)$-point conformal  block in the comb channel. The primary operators are in $\{z_{\infty}, z_1,..., z_n, z_0\}$; the external dimensions are $\{h_\infty, h_1, ... , h_n, h_0\}$ and the intermediate dimensions are $\{\Delta_2, ... , \Delta_{n-3}\}$.}
\label{fig:comb}
\end{figure}

The $(n+2)$-point CPW with external  dimensions $\{h_j, h_0, h_{\infty},\, j=1,...,n\}$, and intermediate dimensions $\{\dl_i,\, i=2,...,n\}$  is defined as
\cite{Rosenhaus:2018zqn}\footnote{This way of labelling coordinates and dimensions is handy when dealing with torus CPWs. As we will see later, the $n$-point torus CPW can be  expressed through the $(n+2)$-point plane CPWs.}
\begin{multline}
\label{planecpw}
\Psi_{\dl_2,\ldots,\dl_{n}}^{h_{\infty}, h_1,\ldots,h_n, h_0}(z_{\infty},z_1,z_2,... ,z_n, z_0) =
\\
= \int_{\mathbb{R}^{n-1}} \left[  {\rm d} w \right]_2^n\, V_{h_{\infty},h_1,\dl_2}(z_{\infty},z_1,w_2) V_{1-\dl_2 ,h_2, \dl_3}(w_2,z_2,w_3) 
\,... \,  V_{1-\dl_{n},h_{n},h_0}(w_n,z_{n},z_0) \,,
\end{multline}
where the measure is defined to be
\be
\label{measure}
\left[  {\rm d} w \right]_i^j \equiv  \prod_{k=i}^{j}  {\rm d} w_k \,.
\ee
This CPW is a linear combination of $2^{n-1}$ terms: a conformal block in  the comb channel (see Fig. \bref{fig:comb}) plus shadow blocks obtained by dualizing the intermediate dimensions $\dl_i \rightarrow 1-\dl_i$, $i=2,..., n$ in all possible ways. Since the comb  diagrams with $n+1$ and $n+2$ endpoints are related  by gluing additionally a 3-point function a remarkable observation was made in  \cite{Rosenhaus:2018zqn} that a given  $(n+2)$-point CPW for $n\geq 3$ can be recursively represented in terms of $(n+1)$-point CPWs as
\begin{multline}
\label{planecpwrecurs}
\Psi_{\dl_2,\ldots,\dl_{n}}^{h_{\infty}, h_1,...\,,h_n, h_0}(z_{\infty},z_1,z_2,...\, ,z_n, z_0) =\\
=\int_{\mathbb{R}} {\rm d} w_n\, \Psi_{\dl_2,...,\dl_{n-1}}^{h_{\infty}, h_1,\ldots,h_{n-1}, \dl_n}(z_{\infty},z_1,z_2,...\, ,z_{n-1}, w_n) V_{1-\dl_{n},h_n,h_0}(w_n,z_n,z_0) \,.
\end{multline}
It is this form of CPWs which finally allows finding closed-form formulas for multipoint conformal blocks. Since the $(n+1)$-point CPW in the integrand (the first factor) is a sum of a conformal block and its shadows (for $\dl_i$, $i=2,..., n-1$), then one can consider only the conformal block contribution by choosing an appropriate integration domain. This restricts  the above recursion relation to conformal blocks only. By a straightforward algebra one finds the $(n+2)$-point global conformal block which is expressed in terms of the comb function \cite{Rosenhaus:2018zqn} (see Appendix \ref{app:multiblocks}).\footnote{See also \cite{Fortin:2019zkm,Fortin:2020zxw,Anous:2020vtw} for the conformal blocks in the comb and  other channels.}

\section{Global torus conformal blocks}
\label{sec:TorusBl}

\begin{figure}
\center{\includegraphics[width=0.4\linewidth]{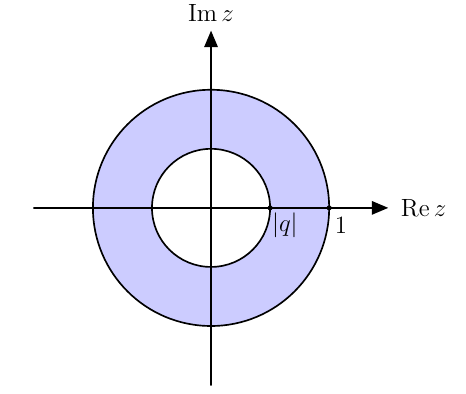}}
\caption{The torus as an annulus on a complex $z$-plane with identified boundary circles   \eqref{zqz1}. The modular parameter $|q| \in (0,1)$.}
\label{fig:annulus}
\end{figure}

Let us begin by  briefly reviewing  the basic structures from torus CFT$_2$ (for a more complete discussion see \cite{Alkalaev:2022kal}).  A two-dimensional torus $\mathbb{T}^2$ can be viewed as a cylinder of the height $\operatorname{Im}\tau$, its circumference equals one, the boundaries are identified with a proper twist $\operatorname{Re}\tau$. The modulus $\tau$ takes values in the fundamental domain $\subset$ the upper half-plane $\mathbb{H}$. In local coordinates on the real plane $t,\bar{t}\in \mathbb{R}^2$ the cylinder is then realised as $t\sim t+1$ and $t\sim t+\tau$. By  $z={e}^{2\pi i t}$ the cylinder can be mapped onto the  complex plane $z,\bar z \in \mathbb{C}$ with  $z\sim q z$, where $q={e}^{2\pi i \tau}$, $q \bar{q} < 1$ is the  modular parameter. Using the latter description  all calculations in torus \cfttwo can be reduced to those  in  plane CFT$_2$, keeping in mind that the coordinates are subject to the equivalence relation, 
\be
\label{zqz}
z\sim q z\,.
\ee
More practically, this  identification can be resolved as a finite domain of $z$-coordinates, namely,\footnote{\label{fn7}Note that $z \bar{z}=\operatorname{e}^{-4\pi \im t}> 0$. Then, by  acting repeatedly   with $z \sim z q$ the plane  $\mathbb{R}^2$ can be transformed to the  annulus  $q \bar{q} \leq z \bar{z} \leq 1$ and  $0 < q \bar{q} <1$ with the identified boundaries,  $1 \sim q \bar{q}$.}
\be
\label{zqz1}
q \bar{q} \leq z \bar{z} \leq 1\,.
\ee
Its boundaries   are identified since $1\sim q$ due to \eqref{zqz}, see Fig. \bref{fig:annulus}. Note that a coordinate  transformation that moves $z$ out of the domain \eqref{zqz1}  must be accompanied by some modular transformation of $q$ which properly changes  the inner boundary $q\bar q$ (and vice versa), thereby guaranteeing that we stay on the same torus.

Let us introduce the group-theoretic notation. Global conformal symmetries in \cfttwo are governed by $sl(2,\mathbb{C})$ with commutation relations $[L_m,L_n]=(m-n)L_{m+n}$ (plus the same for $\bar{L}_m$), where $m,n=0,1,-1$.  The (chiral) Verma module  $\cV_h$ is spanned by the basis vectors (at level $m$) $|m,h\rangle = L_{-1}^m |h\rangle$, with $|h\rangle$ being the highest-weight vector, $L_0 |h\rangle = h |h\rangle$. The Gram matrix in $\cV_h$ is denoted by $\left( B_h \right)^{mn} = \langle h, m| n,h\rangle = \delta_{mn} m! (2 h)_m$, where $(a)_m = \Gamma(a+m)/\Gamma(a)$ is the Pochhammer symbol.

\subsection{Torus correlation functions}
\label{sec:BasicsTorus}


Torus correlation functions of $n$ primary operators $\phi_i(z_i,\zb_i)$ of the (anti)holomorphic  dimensions $h_i,\bar{h}_i$  can be written as
\be
\label{TorusCorr}
\big\langle Y(\bz,\bar{\bz}) \big\rangle_{\tau} = \operatorname{Tr}_\cH \left[Y(\bz,\bar{\bz}) \, q^{L_0} \bar{q}^{\bar{L}_0}\right].
\ee
Here, we denoted  $Y(\bz,\bar{\bz}) \equiv \phi_1(z_1,\zb_1)...\, \phi_n(z_n,\zb_n)$ and $\bz = \{z_1,...\,,z_n \}$ are  points on $\mathbb{C}$, 
the trace is taken over the  Hilbert space of states  
\be
\cH = \bigoplus_{h,\hb \in D}\cV_h \otimes \bar{\cV}_{\bar{h}}\,,
\ee 
where the domain $D$ of admissible dimensions  is determined by choosing  a particular CFT$_2$. Expanding  the trace in \eqref{TorusCorr} yields  
\be
\label{calctrace}
\big\langle Y(\bz,\bar{\bz}) \big\rangle_{\tau} = \sum_{\tilde{h},\bar{\tilde{h}}\in D} \sum_{m,\bar{m}=0}^\infty \left(B_{\tilde{h}}^{-1}\right)^{mm} \left(B_{\bar{\tilde{h}}}^{-1}\right)^{\bar{m}\bar{m}}
\langle \tilde{h},\bar{\tilde{h}},m,\bar{m}|Y(\bz,\bar{\bz}) \,q^{L_0} \bar{q}^{\bar{L}_0} |m,\bar{m},\tilde{h},\bar{\tilde{h}} \rangle \,.
\ee
Thus, the $n$-point torus correlation function is expressed through the $(n+2)$-point plane matrix elements. Since $L_0|m,h\rangle=(m+h)|m,h\rangle$, then  the torus correlation function \eqref{calctrace} is a power series in $q$. At  $q \rightarrow 0$ (when the torus decompactifies onto a plane) all higher-level matrix elements in  \eqref{calctrace} are suppressed and the torus $n$-point correlation function becomes equal to the plane $(n+2)$-point correlation function (see \cite{Alkalaev:2022kal} for a detailed discussion). When considering the torus conformal block as a part of the torus CPW  we will treat this behaviour as the asymptotic  condition: 
\be
\label{asympt}
\cF_{\dl_1,\dl_2,..., \dl_n}^{h_1,h_2,...,h_n}(q \rightarrow 0,z_1,z_2,...\,, z_n) \,\sim\, q^{\Delta_1} G^{\dl_1,h_1,...,h_n,\dl_1}_{\dl_2,...,\dl_n}(\infty, z_1,...\,, z_n, 0) \,, 
\ee
where $\cF$ is the $n$-point torus conformal block, $G$ is the $(n+2)$-point plane block and $\dl_1 \in D$ is the torus intermediate dimension coming from the trace (see the next section). 

The torus geometry breaks the global conformal symmetry of the plane such that the global torus symmetry is only translational,  $u(1)\oplus u(1) \subset sl(2,\mathbb{C})$, generated by $L_0,\bar{L}_0$ (in the plane parametrization these transformations become dilatations). It constrains the  torus correlation functions by implementing the global Ward identities. The holomorphic Ward identity reads as
\be
\label{wardtorus}
\sum_{i=1}^n \cL_0^{(i)}\big\langle \phi_1(z_1,\zb_1) \ldots \phi_i (z_i,\zb_i) \ldots \phi_n(z_n,\zb_n) \big\rangle_{\tau} = 0 \,,
\ee
where $\cL_0$ is an action of $L_0$ on  primary operators (the superscript means acting on $i$-th coordinate)
\be
\label{cL}
\left[ L_m, \phi_{h,\bar{h}}(z,\zb) \right] = \cL_m \phi(z,\zb), \qquad \cL_m = z^m \left(z \partial_z + h(m+1) \right) \,, \quad m = 0, \pm 1\,,
\ee
along with  the same relations for $\bar{L}_m$.

\subsection{Torus shadow formalism}
\label{sec:toruscpw}

In this section we define a torus CPW within the shadow formalism which will be the first step towards the calculation of torus conformal blocks.\footnote{Conformal partial waves in thermal \cft on $S^{1}\times S^{d-1}$  were  discussed in \cite{Gobeil:2018fzy} in the context of the AdS-integral representation of the thermal 1-point block. For more discussion about the relation between the shadow formalism and the geodesic Witten diagrams see \cite{Czech:2016xec,Dyer:2017zef,Chen:2017yia}. } To this end, consider first 1-point torus functions.  The respective CPW can be obtained by inserting the projector \eqref{projector}  in the correlation function \eqref{calctrace} between the left states and $Y(\bz,\bar{\bz}) = \phi(z_1,\zb_1)$: 
\begin{multline}
\label{insertedproj}
\sum_{\tilde{h},\bar{\tilde{h}}\in D} \sum_{m,\bar{m} = 0}^\infty \int_{\mathbb{R}^2} {\rm d}^2 w \left(B_{\tilde{h}}^{-1}\right)^{m m } \left(B_{\bar{\tilde{h}}}^{-1}\right)^{\bar{m} \bar{m} }  \langle \tilde{h}, \bar{\tilde{h}}, m,\bar{m} |\cO(w,\wb)|0\rangle 
\\
\times \langle 0| \widetilde{\cO}(w,\wb)  \phi(z_1,\zb_1) q^{L_0} \bar{q}^{\bar{L}_0} |m,\bar{m}, \tilde{h},\bar{\tilde{h}} \rangle \,.
\end{multline}
Representing the  primary  operator in any point as 
\be
\label{transoper}
\cO(w,\wb) = e^{wL_{-1}+\wb \bar{L}_{-1}}\, \cO(0,0)\, e^{-(wL_{-1}+\wb \bar{L}_{-1})} \,,
\ee
we  find that it creates from the vacuum an infinite linear combination of descendant states 
\be
\label{primaryexpansion}
\begin{split}
\cO(w,\wb) |0\rangle &= \sum_{n,\bar{n}=0}^{\infty} \frac{w^n}{n!} \frac{\wb^{\bar{n}}}{\bar{n}!}L_{-1}^n \bar{L}_{-1}^{\bar{n}} |\dl,\dlb\rangle= \sum_{n,\bar{n}=0}^{\infty} \frac{w^n}{n!} \frac{\wb^{\bar{n}}}{\bar{n}!} |n,\dl \rangle \otimes |\bar{n},\dlb \rangle \,,
\end{split}
\ee
where we used the conformal invariance of the vacuum state,  $L_{m}|0\rangle=0$ and $\bar{L}_{m}\ket{0}=0$. Then, \eqref{insertedproj}   can be rewritten as (to keep things simple  we suppress  $\zb$-dependence for a while) 
\be
\label{steprocpw}
\ba{c}
\dps
\sum_{\tilde{h}\in D} \sum_{m=0}^\infty \sum_{n=0}^\infty \int {\rm d}w  \frac{w^n}{n!}\left(B_{\tilde{h}}^{-1}\right)^{m m } \langle \tilde{h}, m| n, \dl \rangle \langle \widetilde{\cO}(w) \phi_1(z_1) q^{L_0} |m,\tilde{h} \rangle 
\vspace{2mm}
\\
\dps
\hspace{20mm}= \int {\rm d}w  \big\langle \tilde \cO(w) \phi_1(z_1) q^{L_0} \cO(w) \big\rangle.
\ea
\ee
Here, we used  $\langle \tilde{h}, m| n, \dl \rangle = \delta_{\tilde{h},\dl}\left(B_{\dl}\right)^{mn} = \delta_{\tilde{h},\dl} \delta_{m,n} m! (2 \Delta)_m$ and then by resolving the Kronecker symbols we singled out the terms with $\tilde{h}=\dl$ and $m=n$; after that, using \eqref{primaryexpansion} we  summed up the series for $\cO$. One can go further and expand $q={e}^{2 \pi i \tau}$ in $\tau$: 
\be
\label{qexpansion}
\begin{split}
q^{L_0}\cO(w)|0\rangle &= \sum_{k,n=0}^\infty \frac{(2\pi i \tau\, L_0)^k}{k!} \frac{w^n}{n!}|n,\dl \rangle 
=\sum_{k,n=0}^\infty \frac{(2\pi i \tau(\dl+n))^k}{k!} \frac{w^n}{n!}|n,\dl \rangle \\
&=\sum_{n=0}^\infty q^{\dl+n} \frac{w^n}{n!}|n,\dl\rangle=q^{\dl}\sum_{n=0}^\infty  \frac{(q w)^n}{n!}|n,\dl\rangle = q^{\dl} \cO(q w)|0\rangle \,,
\end{split}
\ee
where we again used   \eqref{primaryexpansion} and $L_0 |n,\dl\rangle = (\dl+n)|n,\dl\rangle$. Now, substituting this relation into  \eqref{steprocpw} one obtains an expression which will be considered as a definition of the $1$-point torus CPW:
\be
\label{toruscpw1full}
\Upsilon_{\dl,\dlb} ^{h_1,\bar{h}_1}(q,\bar{q},z_1,\zb_1) 
= q^{\dl} \bar{q}^{\dlb} \int_{\mathbb{R}^2} {\rm d}^2 w\, V_{1-\dl,h_1,\dl}(w,z_1,q w) \bar{V}_{1-\dlb,\bar{h}_1,\dlb}(\wb,\zb_1, \bar{q}\wb)\,,
\ee
with the 3-point functions $V_{h_i,h_j,h_k}$  given by \eqref{vertex}. 

In order to define an $n$-point torus CPW one follows essentially the same procedure with one more step of inserting additional projectors \eqref{projector} between each pair of primary operators in  $Y(\bz,\bar{\bz})$:
\be
\label{toruscpwfull}
\ba{c}
\dps
\Upsilon_{{\bf \dl}, \bar{{\bf \dl}}}^{{\bf h}, \bar{\bf h}}(q,\bar{q},\bz,\bar{\bz}) 
= q^{\dl_1} \bar{q}^{\dlb_1} \int_{\mathbb{R}^{2n}} \left[  {\rm d} w \right]_1^n \left[  {\rm d} \wb \right]_1^n 
\left(\prod_{j=1}^n V_{1-\dl_j,h_j,\dl_{j+1}}(w_j,z_j,w_{j+1})\right) 
\vspace{2mm}
\\
\dps
\hspace{70mm} \times \left(\prod_{\bar{j}=1}^n \bar{V}_{1-\dlb_{\bar{j}},\hb_{\bar{j}},\dlb_{\bar{j}+1}}(\wb_{\bar{j}},\zb_{\bar{j}},\wb_{\bar{j}+1})\right).
\ea
\ee
Here, the antiholomorphic measure is obtained from \eqref{measure} by substituting $w\rightarrow \wb$, we denoted ${\bf \dl} = \{ \dl_1,\ldots \dl_n \}$, ${\bf h} = \{ h_1,\ldots, h_n\}$, and introduced identifications $\dl_{n+1}\equiv \dl_1$ and $w_{n+1} \equiv q w_1$ that guarantees a non-trivial $q$-dependence of the RHS, cf. \eqref{toruscpw1full} (+ the same in the  antiholomorphic sector). In  torus \cfttwo, the procedure of  inserting  projectors and/or using  OPEs leads to  topologically inequivalent channels.\footnote{This is in contrast to  plane CFT$_{2}$. For a detailed discussion see \cite{Alkalaev:2022kal}.} A channel obtained  by inserting the projectors only (as in \eqref{toruscpwfull}) is called the necklace channel. In the rest of the paper we consider torus conformal blocks only in the necklace channel.

Remarkably, the  torus CPW \eqref{toruscpwfull} can be represented in terms of  the plane CPW \eqref{planecpw} as follows: 
\be
\label{toruscpwplanefull}
\ba{l}
\dps
\Upsilon_{{\bf \dl}, \bar{{\bf \dl}}}^{{\bf h}, \bar{\bf h}}(q,\bar{q},\bz,\bar{\bz}) =
\vspace{2mm}
\\
\dps 
\hspace{7mm}= q^{\dl_1} \bar{q}^{\dlb_1} \int_{\mathbb{R}^2} {\rm d}^2 w \; \Psi^{1-\dl_1,h_1,\ldots,h_n,\dl_1; 1-\dlb_1, \bar{h}_1,\ldots, \bar{h}_n,\dlb_1}_{\dl_2,\ldots,\dl_n; \dlb_2,\ldots, \dlb_n} (w,\wb,z_1,\zb_1,...\,, z_n, \zb_n, q w, \bar{q} \wb) \,.
\ea
\ee
Indeed, substituting \eqref{planecpw} into \eqref{toruscpwplanefull} results in \eqref{toruscpwfull}. It is not occasional because the $n$-point torus correlation function  is expressed in terms of the $(n+2)$-point plane matrix elements \eqref{calctrace}.


Having in mind the (anti-)chiral factorization of \cfttwo torus conformal blocks, it is convenient  to introduce a chiral CPW: 
\be
\label{toruscpwvert}
\Upsilon_{{\bf \dl}}^{{\bf h}}(q,\bz) 
= q^{\dl_1} \int_{\mathbb{R}^n} \left[  {\rm d} w \right]_1^n\,
\Bigg(\prod_{j=1}^n V_{1-\dl_j,h_j,\dl_{j+1}}(w_j,z_j,w_{j+1})\Bigg) \,,
\ee
where, as before, $\dl_{n+1}\equiv \dl_1$ and $w_{n+1} \equiv q w_1$. From  \eqref{planecpw} one derives an equivalent form 
\be
\label{toruscpwplane}
\Upsilon_{{\bf \dl}}^{{\bf h}}(q,\bz) = q^{\dl_1}\int_{\mathbb{R}} {\rm d} w\; \Psi_{\dl_2,\ldots, \dl_n}^{1-\dl_1,h_1,\ldots,h_n,\dl_1}(w,z_1,...\,,z_n, q w) \,.
\ee

From now on we consider the coordinates and the modular parameter to be real, $z_i \in \mathbb{R}$,  $0<q<1$. For real coordinates the identification \eqref{zqz} keeps its form,  $z\sim q z$, which can be now resolved as 
\be
\label{zqz2}
q \leq z \leq 1\,.
\ee 
Similarly  to \eqref{zqz1} the boundary points here are  identified as $1 \sim q$ that results in $z \in S^1$. Indeed,  the real  modular parameter $q$ (i.e. $\re \tau = 0$) corresponds to a rectangular torus $T^2 = S^1 \times S^1$ where the first and second circles have radii $1$ and $\im \tau$, respectively. Then, using the cylinder map $z = \exp (2 \pi i \re t - 2 \pi \im t)$ and imposing the reality condition on $z$  one finds that the periodic coordinate identification $t \sim t + 1$ is not valid anymore  since  $\re t = 0$. It follows that the first circle in $T^2 = S^1 \times S^1$ shrinks to a point  and one is left with the second circle of radius $\im \tau$ realized as \eqref{zqz2}. This justifies a somewhat sloppy term  ``a \cftone torus CPW'' which will be used in the sequel to designate one-dimensional CPW on a circle.  

Note that considering the global  transformations $u(1)$ one can show that the \cftone torus CPW \eqref{toruscpwvert} has a scaling dimension $h_1+\ldots +h_n$. It is obvious since the torus CPW is obtained from the torus correlation function \eqref{TorusCorr} upon inserting dimensionless projectors \eqref{projector}. As always, the global invariance can be used to fix one of $z_i$.

\section{$n$-point  torus CPW}
\label{sec:calc}

As discussed  in Section \bref{sec:global}, the plane CPW $\Psi$ can be represented as  a sum of the conformal block and its shadows. Since the torus CPW $\Upsilon$ is expressed through the plane CPWs $\Psi$ by the means of \eqref{toruscpwplanefull} one concludes that $\Upsilon$ can also be represented as a sum of the conformal block and its shadows. In what follows, we explicitly calculate 1-point and 2-point conformal blocks thereby showing in detail the procedure of extracting conformal blocks from CPWs. Then, we generalize this consideration to the $n$-point case.

\subsection{1-point torus block}
\label{1pointTorusBlock}

Let us consider first the $1$-point torus CPW \eqref{toruscpwvert}\footnote{This 1-point example illustrates the invariance condition discussed in the end of the previous section. Indeed, the global transformation $w\rightarrow \lambda w$ changes singularities of the integrand as $w = (0,z/q,z)\rightarrow (0,z/\lambda q, z/\lambda)$. As a result,  $\Upsilon_{\dl}^{h}(q,z) = \lambda^{-h} \Upsilon_{\dl}^{h}(q,z/\lambda)$.}
\be
\Upsilon_{\dl}^{h}(q,z) = q^{\dl} \int_{\mathbb{R}} {\rm d}w\, V_{1-\dl,h,\dl}(w,z,q w) \,.
\ee
By substituting 3-point function \eqref{vertex} and changing the integration variable $w\rightarrow z w$ this expression can be cast into the form
\be
\label{Torus1CPWexplicitly}
\Upsilon_{\dl}^{h}(q,z) = \frac{q^{1-h-\dl} \, z^{-h}}{(1-q)^{1-h}} \, \int_{\mathbb{R}}  \frac{{\rm d}w}{|1-w|^{1+h-2\dl} |1/q-w|^{h+2\dl -1} |w|^{1-h} } \,.
\ee
The integral here is similar to that one which defines the 4-point CPW in  planar \cfttwo \eqref{cpwcross4}. Likewise, the integral is singular, as the integrand has five singular points $w=\{-\infty, 0,1,q^{-1},+\infty \}$. It can be calculated  by splitting the integration domain into subregions whose boundaries are the singular points, cf. \eqref{4pointSplit}. The integrals in each subregion are given by  linear combinations of two hypergeometric functions ${}_2 F_{1}$ that  can be seen from \eqref{cpw4integrals}. Thus, the 1-point torus CPW \eqref{Torus1CPWexplicitly} is a sum of two terms related by  $\dl \rightarrow 1-\dl$. The term satisfying the required  asymptotic condition  \eqref{asympt} is the 1-point torus block given by 
\be
\label{TorusBlock1}
\cF_{\dl}^{h}(q,z)=\Lambda^{h}(q,z) \cV_{\dl}^{h}(q) \,,
\ee
where 
\be
\label{TorusLeg1}
\Lambda^{h}(q,z) = \frac{z^{-h}}{(1-q)^{1-h}}
\ee
 is the torus $1$-point leg factor and 
\be
\label{TorusBareBlock1}
\cV_{\dl}^{h}(q)=q^{\dl} \,{}_2 F_1 
\left[
\begin{array}{l l}
h, h+2 \dl -1 \\
\qquad  2\dl
\end{array}\bigg| \; q
\right] 
\ee
is the global  1-point torus conformal block \cite{Hadasz:2009db}. Note that the hypergeometric series  in \eqref{TorusBareBlock1} converges in the prescribed domain   $q\in (0,1)$. The torus leg factor \eqref{TorusLeg1} depends only on external  dimension $h$ and provides  the covariance  of the conformal block \eqref{TorusBlock1} under conformal transformations. It has the same dimension as the torus $1$-point function.

At $h=0$ the 1-point torus block \eqref{1pointTorusBlock} becomes equal to $\cF^{h=0}_{\dl}(q,z) = q^{\dl}/(1-q)$ which is the $sl(2, \mathbb{R})$ character of the Verma module $\cV_\dl$. The same can be directly obtained from evaluating the $0$-point \cfttwo torus CPW\footnote{Its definition follows from \eqref{steprocpw} after choosing  $\phi(z, \zb) = \mathbb{1}$, i.e. $h_1=\bar{h}_1=0$ and then following the same steps as in \eqref{qexpansion}.}
\be
\label{zerocpw}
\Upsilon_{\dl,\dlb}(q,\bar{q}) =q^{\dl} \bar{q}^{\dlb} \int_{\mathbb{R}^2} {\rm d}^2 w \langle \widetilde{\cO}(w,\wb) \cO(q w, \bar{q} \wb) \rangle \,.
\ee
Using the 2-point function for the equivalent representations \eqref{shadow2pt}  and the standard formula for the $\delta$-function (see the footnote \bref{fn1}) one obtains 
\be
\Upsilon_{\dl,\dlb}(q,\bar{q})= \frac{q^{\dl}}{(1-q)}\frac{\bar q^{\bar{\dl}}}{(1-\bar q)} \,,
\ee
which is the $sl(2, \mathbb{C})$ character of the Verma module $\cV_\dl \otimes \cV_{\bar \dl}$. Thus, we  conclude that the definition of the torus CPW \eqref{toruscpwvert}-\eqref{toruscpwplane} works well for the known examples. 

\subsection{2-point torus block}
\label{sec:torus2}

Just as in the case of the plane blocks,  we will only focus on the functional dependence and we will not be interested in the overall coefficients  depending  on external/intermediate  dimensions. 

For $n=2$ the \cftone torus CPW \eqref{toruscpwplane} reads as 
\be
\label{toruscpw2}
\Upsilon_{\dl_1,\dl_2}^{h_1,h_2}(q,z_1,z_2) = q^{\dl_1} \int_{\mathbb{R}} {\rm d}w\, \Psi_{\dl_2}^{1-\dl_1,h_1,h_2,\dl_1}(w,z_1,z_2,q w) \,.
\ee
Here, the integral is given by the  sum of 4 terms: the torus conformal block plus its  shadows obtained  by $\dl_1 \rightarrow 1-\dl_1$ and $\dl_2 \rightarrow 1-\dl_2$. 

The plane 4-point CPW $\Psi$ in the integrand of \eqref{toruscpw2} is a sum of the plane conformal block and its shadow \eqref{cpw4sum}. To avoid shadow block terms we consider only the 4-point conformal block contribution to the plane CPW\footnote{Notation $\mbox{LHS} \supset \mbox{RHS}$ introduced in \cite{Rosenhaus:2018zqn} means that the  RHS is one of the terms contained on the LHS, moreover, a $(\Delta,h)$-dependent prefactor on the RHS is dropped.} 
\be
\Upsilon_{\dl_1,\dl_2}^{h_1,h_2}(q,z_1,z_2) \supset q^{\dl_1} \int_{\mathbb{R}}  {\rm d}w \, G_{\dl_2}^{1-\dl_1,h_1,h_2,\dl_1} 
(w,z_1,z_2,q w)\,,
\ee
where the 4-point plane block is given by \eqref{block4} (expanded as the hypergeometric series)
\begin{multline}
G_{\dl_2}^{1-\dl_1,h_1,h_2,\dl_1} (w,z_1,z_2,q w)= \cL^{1-\dl_1,h_1,h_2,\dl_1} (w,z_1,z_2,q w)\\
\times \, \sum_{m=0}^{\infty}\frac{(1-\dl_1+\dl_2-h_1)_m (\dl_1+\dl_2-h_2)_m}{(2 \dl_2)_m m!} \left(\frac{(w-z_1) \left(\frac{z_2}{q}-w \right)}{(w-z_2)\left(\frac{z_1}{q}-w \right)}\right)^{m+\dl_2},
\end{multline}
with the leg factor \eqref{leg4}
\be
\cL^{1-\dl_1,h_1,h_2,\dl_1} = \frac{\;(w-z_1)^{\dl_1-h_1-1}}{ (z_2-q w)^{h_2+\dl_1}}\left(\frac{z_{12}}{w-z_2}\right)^{1-\dl_1-h_1} \left(\frac{z_1-q w}{z_{12}}\right)^{h_2-\dl_1}.
\ee
Thus, we have  the 2-point torus CPW in the form
\begin{multline}
\Upsilon_{\dl_1,\dl_2}^{h_1,h_2}(q,z_1,z_2) \supset q^{-\dl_1} z_{12}^{1-h_1-h_2} \sum_{m=0}^\infty \frac{(1-\dl_1+\dl_2-h_1)_m (\dl_1+\dl_2-h_2)_m}{(2 \dl_2)_m m!} 
\\
\times \int_{\mathbb{R}}  {\rm d}w\; \frac{(w-z_1)^{\dl_2+m-h_1+\dl_1-1}}{ (w-z_2)^{\dl_2+m-h_1-\dl_1+1}}\, \left(\frac{z_1}{q}-w \right)^{h_2-\dl_2-m-\dl_1} \left(\frac{z_2}{q}-w \right)^{-h_2+\dl_2+m-\dl_1}.
\end{multline}
The integral in the second line is recognized as the 4-point CPW \eqref{cpw4final}:
\be
\label{cpwtor2cpw4contr}
\ba{l}
\dps
\Upsilon_{\dl_1,\dl_2}^{h_1,h_2}(q,z_1,z_2) \supset \sum_{m=0}^\infty (-)^m\,\frac{(1-\dl_1+\dl_2-h_1)_m (\dl_1+\dl_2-h_2)_m}{(2 \dl_2)_m \,m!} 
\vspace{2mm}
\\
\dps
\hspace{33mm} \times \, q^{-\dl_2-m-h_2} z_{12}^{2m+2\dl_2} \, \Psi_{\dl_1}^{\dl_2+m,h_2,h_1,\dl_2+m}\left(\frac{z_1}{q},\frac{z_2}{q}, z_1,z_2 \right).
\ea
\ee

As in plane \cfttwo there is no need  to integrate over all $\mathbb{R}$ to find the 2-point necklace block. By choosing an appropriate integration domain in $\Psi_{\dl_1}^{\dl_2+m,h_2,h_1,\dl_2+m}$ one can single out the conformal block contribution $G_{\dl_1}^{\dl_2+m,h_2,h_1,\dl_2+m}$. In the present case, the integration domain is split into four intervals by five singular points $\{-\infty,0,\rho_1,1,+\infty \}$, where  
\be
\label{rho_ratio1}
\rho_1 = \frac{q (z_{12})^2}{z_1 z_2 (1-q)^2 } = \frac{q (1-x_1)^2}{x_1 (1-q)^2}\,, \qquad x_1 = \frac{z_2}{z_1} \,,
\ee
is just the plane cross-ratio $\chi_1$ evaluated with respect to four arguments of $\Psi_{\dl_1}^{\dl_2+m,h_2,h_1,\dl_2+m}$ in \eqref{cpwtor2cpw4contr}. Note that $x_1$ is the $u(1)$ invariant. Then, choosing the second interval $(0, \rho_1)$  one is left with\footnote{See \eqref{cpw4integrals} and the term proportional $a_2$. In the torus case, however, the resulting $a_2$  depends on the  summation index $m$ (the first line in \eqref{cpw4sumindneck2}). Thus, contrary to the plane case, it cannot be dropped.} 
\begin{multline}
\label{cpw4sumindneck2}
\Psi_{\dl_1}^{\dl_2+m,h_2,h_1,\dl_2+m} \supset \frac{\Gamma(\dl_1+h_1-\dl_2-m) \Gamma(\dl_1+\dl_2+m-h_1)}{\Gamma(2\dl_1)}\\
\times \, \mathcal{L}^{\dl_2+m,h_2,h_1,\dl_2+m}\left(\frac{z_1}{q},\frac{z_2}{q}, z_1,z_2 \right) g_{\dl_1}^{\dl_2+m,h_2,h_1,\dl_2+m}(\rho_1) \,,
\end{multline}
where the leg factor which is  \eqref{leg4} in the new points is given by 
\be
\mathcal{L}^{\dl_2+m,h_2,h_1,\dl_2+m} = \frac{q^{\dl_2+m+h_2}}{(z_{12})^{h_1+h_2+2\dl_2+2m}} \left(\frac{z_2-q z_1}{z_1(1-q)}\right)^{\dl_2+m-h_2} \left( \frac{z_2(1-q)}{z_2-q z_1}\right)^{h_1-\dl_2-m}\,,
\ee
and the bare block \eqref{blockbare4} expanded into a series reads
\be
\label{bareb}
g_{\dl_1}^{\dl_2+m,h_2,h_1,\dl_2+m}(\rho_1) =   \sum_{n=0}^{\infty} \frac{(\dl_1+\dl_2+m-h_2)_n (\dl_1+\dl_2+m-h_1)_n }{(2\dl_1)_n}\rho_1^{\dl_1+n}\,.
\ee 
Using the identities $\Gamma(1-s)\Gamma(s)=\pi/\sin (\pi s)$ and $(a)_m = \Gamma(a+m)/\Gamma(a)$ the prefactor in \eqref{cpw4sumindneck2} can be written as 
\be
\label{Gamma_id}
\begin{split}
\Gamma(\dl_1+h_1-\dl_2-m)& \Gamma(\dl_1+\dl_2+m-h_1) \\
=& \frac{\pi \, \Gamma(\dl_1+\dl_2+m-h_1)}{\sin (\pi(\dl_1+h_1-\dl_2-m)) \,\Gamma(1-\dl_1-h_1+\dl_2+m)} \\
=& \frac{\pi \,(-)^m}{\sin (\pi(\dl_1+h_1-\dl_2))}\frac{(\dl_1+\dl_2-h_1)_m \, \Gamma(\dl_1+\dl_2-h_1)}{(1-\dl_1+\dl_2-h_1)_m \, \Gamma(1-\dl_1+\dl_2-h_1)}\,.
\end{split}
\ee
Thus, substituting \eqref{Gamma_id} and \eqref{bareb} back into \eqref{cpw4sumindneck2} along with using $(a+m)_n = (a)_{m+n}/(a)_m$   yields   
\be
\label{cpw4sumindneck2final}
\ba{l}
\dps
\Psi_{\dl_1}^{\dl_2+m,h_2,h_1,\dl_2+m} \supset \frac{(-)^m}{(1-\dl_1+\dl_2-h_1)_m (\dl_1+\dl_2-h_2)_m} \mathcal{L}^{\dl_2+m,h_2,h_1,\dl_2+m}\left(\frac{z_1}{q},\frac{z_2}{q}, z_1,z_2 \right)
\vspace{2mm} 
\\
\dps 
 \hspace{45mm} \times\, \sum_{n=0}^{\infty}\frac{(\dl_1+\dl_2-h_2)_{m+n} (\dl_1+\dl_2-h_1)_{m+n} }{(2\dl_1)_n}\rho_1^{\dl_1+n} \,.
\ea
\ee
Plugging \eqref{cpw4sumindneck2final} into \eqref{cpwtor2cpw4contr} one finally gets  
\be
\Upsilon_{\dl_1,\dl_2}^{h_1,h_2}(q,z_1,z_2) \supset \cF_{\dl_1,\dl_2}^{h_1,h_2}(q,z_1,z_2) \,,
\ee
where $\cF$ is the 2-point torus conformal block in the necklace channel 
\be
\label{torusblock2}
\cF_{\dl_1,\dl_2}^{h_1,h_2}(q,z_1,z_2) = \Lambda^{h_1,h_2}(q,z_1,z_2)\, \cV_{\dl_1,\dl_2}^{h_1,h_2}(\rho_1,\rho_2) \,,
\ee
with the 2-point torus leg factor
\be
\label{torusleg2}
\Lambda^{h_1,h_2}(q,z_1,z_2) = \frac{z_1^{h_2} \,z_2^{h_1} \,(1-q)^{h_1+h_2}}{(z_{12})^{h_1+h_2}\, (z_2-q z_1)^{h_1+h_2}} \,,
\ee
and the 2-point bare torus block
\be
\label{torusbare2}
\cV_{\dl_1,\dl_2}^{h_1,h_2}(\rho_1,\rho_2) = \rho_1^{\dl_1} \rho_2^{\dl_2}\, F_4 \left[
\begin{array}{l l}
\dl_1+\dl_2 -h_1, \dl_2+\dl_1-h_2 \\
\qquad  \qquad 2\dl_1, 2\dl_2
\end{array}\bigg| \rho_1 , \rho_2
\right],
\ee
where $F_4$ is the fourth Appell function \eqref{appelfunction4} and 
\be
\label{rho_ratio2}
\rho_2 = \frac{(z_2-q z_1)^2}{z_1 z_2 (1-q)^2} = \frac{(x_1-q)^2}{x_1 (1-q)^2} \,,
\ee
where $x_1$ is defined  in \eqref{rho_ratio1}. Note that $\rho_2$ arises when one substitutes  \eqref{cpw4sumindneck2final} into \eqref{cpwtor2cpw4contr} and then assembles all contributions in $z$ and $q$ of $m$-power into  $\rho_2^{\dl_2+m}$ that defines a second argument in the  double power series. The quantities $\rho_1\in (0,1)$ \eqref{rho_ratio1} and $\rho_2\in(0,1)$  \eqref{rho_ratio2} will be referred to as the torus cross-ratios. The Appell series $F_4$ converges when $\sqrt{\rho_1}+\sqrt{\rho_2} < 1$ \cite{Bateman:100233}. This inequality is satisfied in the prescribed domain of the modular parameter $q \in (0,1)$ and  external points $x_1 \in (q,1)$ (see Section \bref{sec:TorusBl}). 

Two comments are in order. First,  the small-$q$ asymptotics  of  \eqref{torusblock2} is given by 
\be
\label{boundary2pt}
\cF_{\dl_1,\dl_2}^{h_1,h_2}(q \rightarrow 0,z_1,z_2) \sim q^{\dl_1} G_{\dl_2}^{\dl_1,h_1,h_2,\dl_1}(\infty,z_1,z_2,0) \,, 
\ee
where $G_{\dl_2}^{\dl_1,h_1,h_2,\dl_1}$ is a particular  4-point plane block in the comb channel \eqref{block4} that agrees with the requirement  \eqref{asympt}. Second, one can explicitly check that the 2-point torus block solves the torus Casimir equations \cite{Alkalaev:2022kal} (see Appendix \bref{app:B2}).

\paragraph{Torus cross-ratios.} These particular combinations of  $z_i$ and $q$ arise naturally when calculating conformal blocks within the torus shadow formalism. We saw that the torus cross-ratios provide a convenient parametrization which allows representing the conformal block in a closed-form. For the bare block we have  $\cV_{\dl_1,\dl_2}^{h_1,h_2}(x_1,q) \sim F_4(\rho_1,\rho_2)$ that is the original set of  variables $x_1 = z_2/z_1$ and $q$ is replaced by $\rho_1(x_1, q)$ and $\rho_2(x_1,q)$, where the $z$-dependence is now spread out  over two variables in globally-invariant way. The standard approaches, e.g. by representing the torus block as a power series in $q$ and calculating the expansion coefficients being the matrix elements of  primary operators, leave no chance to see that $z_i$ and $q$  can be organized into the rational functions because that will require multiple non-trivial  resummations.     

Although the geometrical meaning of the torus cross-ratios is yet to be understood, here we notice  that they are invariant under the following transformations ($\forall \lambda \in \mathbb{R}$):
\be
\label{t1}
z_i\rightarrow \lambda z_i\,, \qquad q \rightarrow q\,; 
\ee
\be
\label{t2}
\hspace{4mm}z_i \rightarrow z^{-1}_i, \qquad q \rightarrow q^{-1}\,.
\ee
A few comments are in order. First, in terms of the modulus $\tau$ the transformations of $q$ in \eqref{t1} and \eqref{t2}  correspond to $\tau \rightarrow \pm \tau$ which  is the identical modular transformation of $PSL(2, \mathbb{Z})$. Second, in terms of the $u(1)$-invariant ratio $x_1 = z_2/z_1$ the transformation \eqref{t1} trivializes, i.e. $x_1 \to x_1$ and $q \to q$. In particular, the torus cross-ratios are $u(1)$ invariants. Third, the transformation \eqref{t1} with the parameter $\lambda = q$ is actually the equivalence relation \eqref{zqz}, while the transformation \eqref{t2} is an inversion of $z_i$ which, therefore, requires a proper modular transformation of $q$ that aims to map the domain of $z_i$ back to \eqref{zqz2} (or, in two dimensions, to \eqref{zqz1}).

\paragraph{Reduction to the 1-point torus block.} Setting  $h_2=0$ and $\dl_1=\dl_2$ one expects that the 2-point necklace block \eqref{torusblock2} is reduced to the 1-point torus block \eqref{TorusBlock1}. Indeed, the 2-point torus leg factor \eqref{torusleg2} takes form 
\be
\label{toursleg2cond}
\Lambda^{h_1,h_2=0}(q,z_1,z_2) = \Lambda^{h_1}(q,z_1) (1-q) \left(\frac{x_1}{(1-x_1) (x_1-q)}\right)^{h_1},
\ee
where $\Lambda^{h_1}(q,z_1)$ is the 1-point torus leg factor \eqref{TorusLeg1},  while the 2-point bare necklace block \eqref{torusbare2} is given by 
\be
\label{torusbare2cond}
\cV^{h_1,h_2=0}_{\dl_1,\dl_1}(\rho_1,\rho_2) = \frac{q^{\dl_1}}{1-q} \left(\frac{(1-x_1) (x_1-q)}{x_1}\right)^{h_1} {}_2 F_1 \left[
\begin{array}{l l}
h_1, 2\dl_1 + h_1-1 \\
\qquad 2\dl_1
\end{array}\bigg| q \right] \,,
\ee
where we used the following identity for the fourth Appell function \cite{Bateman:100233}
\be
\begin{split}
\label{appell4identityred}
F_4 \left[
\begin{array}{l l}
2\dl_1-h_1, 2\dl_1 \\
\quad 2\dl_1, 2\dl_1
\end{array}\bigg| \, \frac{q (1-x_1)^2}{x_1 (1-q)^2},\, \frac{(x_1-q)^2}{x_1 (1-q)^2} \,
\right]& \\
= \frac{1}{1-q} \left(\frac{x_1 \,(1-q)^2}{(x_1-q)(1-x_1)}\right)^{2\dl_1} \left(\frac{(x_1-q)(1-x_1)}{x_1}\right)^{h_1} {}_2 F_1 \left[
\begin{array}{l l}
h_1, 2\dl_1 + h_1-1 \\
\qquad 2\dl_1
\end{array}\bigg| q \right]& \,.
\end{split}
\ee
Thus, as expected, one has
\be
\label{reduction2pt}
\cF_{\dl_1,\dl_1}^{h_1,h_2=0}(q,z_1,z_2) = \cF_{\dl_1}^{h_1}(q,z_1) \,.
\ee
Note that $z_2$ was kept arbitrary, nevertheless, it drops out identically once the constraints $h_2=0$, $\dl_2 = \dl_1$ are imposed.

\subsection{$n$-point torus block}
\label{sec:torusmulti}

Recall that the $n$-point plane CPW $\Psi$  is recursively expressed through the $(n-1)$-point plane CPW  \eqref{planecpwrecurs}.  However, one can see that  the $n$-point torus CPW $\Upsilon$ is not expressed through the $(n-1)$-point torus CPW, but, instead,  through the $(n+2)$-point plane CPW $\Psi$. On the other hand, the recursion relation solves  the $(n+2)$-point plane CPW in terms of  the 4-point plane CPW which implies that the $n$-point torus CPW $\Upsilon$ for any $n$ can be represented as a linear combination of the 4-point plane CPWs. In fact, this phenomenon is seen already in the 2-point case \eqref{cpwtor2cpw4contr} and, therefore, the calculation  of the $n$-point necklace block follows essentially the same steps. 

Let us consider the $(n+2)$-point plane conformal block contribution to the $n$-point torus CPW \eqref{toruscpwplane}
\be
\label{toruscpwblockn}
\Upsilon_{{\bf \dl}}^{{\bf h}}(q,\bz) \supset q^{\dl_1}\int_{\mathbb{R}} {\rm d} w\; G_{\dl_2,\ldots, \dl_n}^{1-\dl_1,h_1,\ldots,h_n,\dl_1}(w,z_1,...\,,z_n, q w) \,,
\ee
where the conformal block is represented as a product of the leg factor and the bare conformal block \eqref{combblock},
\be
\begin{split}
G_{\dl_2,..., \dl_n}^{1-\dl_1,h_1,...,h_n,\dl_1}(w,z_1,...\,,z_n, q w) &= \cL^{1-\dl_1,h_1,...,h_n,\dl_1}(w,z_1,...\,,z_n,q w)\\
&\times \, g_{\dl_2,..., \dl_n}^{1-\dl_1,h_1,...,h_n,\dl_1}(\chi_{\infty},\chi_1,...\,,\chi_{n-3},\chi_0)\,,
\end{split}
\ee
with the leg factor \eqref{combleg},
\begin{multline}
\label{legderivneckn}
\cL^{1-\dl_1,h_1,\ldots,h_n,\dl_1} =\prod_{i=1}^{n-2} \left(\frac{z_{i,i+2}}{z_{i,i+1} \, z_{i+1,i+2}}\right)^{h_{i+1}}\left(\frac{z_{12}}{(w-z_1)(w-z_2)} \right)^{1-\dl_1} \\
\times \, \left( \frac{z_{n-1,n}}{(z_{n-1}-q w) (z_n -q w)}\right)^{\dl_1} \left(\frac{w-z_2}{(w-z_1)\, z_{12}}\right)^{h_1} \left(\frac{z_{n-1}-q w}{z_{n-1,n}\,(z_n - q w)}\right)^{h_n} \,,
\end{multline}
and the bare block \eqref{combbare},
\be
\label{baredervneckn}
g_{\dl_2,\ldots, \dl_n}^{1-\dl_1,h_1,\ldots,h_n,\dl_1} = \sum_{m_2,\ldots,m_n=0}^{\infty}  Y_{m_2,m_3,\ldots,m_n} \,
\chi_{\infty}^{m_2+\dl_2}\chi_{1}^{m_3+\dl_3}\ldots \chi_{n-3}^{m_{n-1}+\dl_{n-1}}\chi_{0}^{m_n+\dl_n} \,,
\ee
where series coefficients of the comb function \eqref{combfunction} are given by
\be
\label{coeffcombblock}
Y_{m_2,m_3,\ldots,m_n} \equiv  \frac{(
1-\dl_1+\dl_2-h_1)_{m_2} (\dl_2+\dl_3-h_2)_{m_2+m_3}\ldots (\dl_n + \dl_1- h_n)_{m_n}}{ (2\dl_2)_{m_2}\ldots (2\dl_n)_{m_n} \, m_2! \, m_3!\, \ldots m_n!}\,,
\ee
and cross-ratios are read off from \eqref{combcross} by setting  $z_{\infty} = w$ and $z_0 = q w$:
\be
\chi_{\infty} = \frac{(w-z_1)\, z_{23}}{(w-z_2)\, z_{13}}\,, \quad \chi_0 = \frac{z_{n-2,n-1} \, (z_n - q w)}{z_{n-2,n}\, (z_{n-1} - q w)}\,, \quad \chi_i = \frac{z_{i,i+1} z_{i+2,i+3} }{z_{i,i+2} z_{i+1,i+3} } \,, \quad 1\leq i \leq n-3 \,.
\ee
Plugging the leg factor \eqref{legderivneckn} and the bare block \eqref{baredervneckn} into  \eqref{toruscpwblockn} one can single out the 4-point CPW \eqref{cpw4final}, namely, 
\begin{multline}
\label{toruscpwblocknfinal}
\Upsilon_{{\bf \dl}}^{{\bf h}}(q,\bz) \supset \sum_{m_2,\ldots,m_n=0}^{\infty} (-)^{m_2} 
\chi_{1}^{m_3+\dl_3}...\, \chi_{n-3}^{m_{n-1}+\dl_{n-1}}\, Y_{m_2,m_3,\ldots,m_n}\, q^{-\dl_1-\dl_n-m_n-h_n}  \\
\times \, \left(\frac{z_{12} \, z_{23}}{z_{13}}\right)^{\dl_2+m_2} \left(\frac{z_{n-2,n-1}\, z_{n-1,n}}{z_{n-2,n}}\right)^{\dl_n+m_n} \Psi_{\dl_1}^{\dl_n+m_n,h_n,h_1,\dl_2+m_2}\left(\frac{z_{n-1}}{q}, \frac{z_n}{q},z_1,z_2 \right),
\end{multline}

As in the  2-point  case, the respective integral can be taken by splitting  the integration domain by five singular points as $\{-\infty,0,\rho_1,1,+\infty \}$, where
\be
\label{rhon_ratio1}
\rho_1 = \frac{q \;z_{12}\; z_{n-1,n}}{(z_{n-1}-q z_1)(z_n-q z_2)} \,.
\ee
which is the plane cross-ratio $\chi_1$ evaluated for the four points which are the arguments 
of $\Psi_{\dl_1}^{\dl_n+m_n,h_n,h_1,\dl_2+m_2}$ in \eqref{toruscpwblocknfinal}. Choosing the interval $(0, \rho_1)$ one gets  
\begin{multline}
\label{cpw4contrneckn}
\Psi_{\dl_1}^{\dl_n+m_n, h_n,h_1,\dl_2+m_2}  \supset \frac{\Gamma(\dl_1+h_1-\dl_2-m_2) \Gamma(\dl_1-h_1+\dl_2+m_2) }{\Gamma(2\dl_1)} \\
\times \, \cL^{\dl_n+m_n, h_n,h_1,\dl_2+m_2} \left( \frac{z_{n-1}}{q},\frac{z_n}{q},z_1,z_2\right) g_{\dl_1}^{\dl_n+m_n, h_n,h_1,\dl_2+m_2}(\rho_1) \,,
\end{multline}
with the leg factor \eqref{leg4}
\be
 \cL^{\dl_n+m_n, h_n,h_1,\dl_2+m_2} = \frac{q^{\dl_n+m_n+h_n}}{z_{n-1,n}^{\dl_n+m_n+h_n} z_{12}^{h_1+m_2+\dl_2}}\left(\frac{z_n-q z_1}{z_{n-1}-q z_1}\right)^{\dl_n+m_n-h_n} \left(\frac{z_n-q z_2}{z_n-q z_1}\right)^{h_1-\dl_2-m_2}\,,
\ee
and the bare block \eqref{blockbare4} expanded as 
\be
g_{\dl_1}^{\dl_n+m_n, h_n,h_1,\dl_2+m_2} = \sum_{m_1=0}^{\infty} \frac{(\dl_1+\dl_n-h_n)_{m_1+m_n} \, (\dl_1+\dl_2-h_1)_{m_1+m_2}}{(\dl_1+\dl_n-h_n)_{m_n}\, (\dl_1+\dl_2-h_1)_{m_2}} \frac{\rho_1^{\dl_1+m_1}}{m_1!\,(2\dl_1)_{m_1}}\,,
\ee
where we  used $(a+m)_n = (a)_{m+n}/(a)_m$. Using the identity \eqref{Gamma_id} (with $m\to m_2$) one finds that  \eqref{cpw4contrneckn} takes the form
\begin{multline}
\label{cpw4contrnecknfinal}
\Psi_{\dl_1}^{\dl_n+m_n, h_n,h_1,\dl_2+m_2}  \supset \frac{(-)^{m_2} \, \cL^{\dl_n+m_n, h_n,h_1,\dl_2+m_2} \left( \frac{z_{n-1}}{q},\frac{z_n}{q},z_1,z_2\right)}{(1-\dl_1+\dl_2-h_1)_{m_2}\,(\dl_1+\dl_n-h_n)_{m_n}} \\
\times \, \sum_{m_1=0}^{\infty} \frac{(\dl_1+\dl_n-h_n)_{m_1+m_n} \, (\dl_1+\dl_2-h_1)_{m_1+m_2}}{(2\dl_1)_{m_1}} \frac{\rho_1^{\dl_1+m_1}}{m_1!}\,.
\end{multline}
Substituting this expression into \eqref{toruscpwblocknfinal} one singles out the necklace torus block contribution to the torus CPW, 
\be
\Upsilon_{{\bf \dl}}^{{\bf h}}(q,\bz) \supset \cF_{{\bf \dl}}^{{\bf h}}(q,\bz) \,.
\ee
Here, the $n$-point block is represented as a product of the torus leg factor and the bare necklace block
\be
\label{necklaceblock}
\cF_{{\bf \dl}}^{{\bf h}}(q,\bz) = \Lambda^{\bf h}(q,\bz) \cV_{\bf \dl}^{\bf h}(\rho_1, ...\,, \rho_n)  \,,
\ee
where the leg factor is defined to be 
\be
\label{necklaceleg}
\Lambda^{\bf h}(q,\bz) =  
\frac{(z_{n-1} - q z_1)^{h_n} (z_n-q z_2)^{h_1}}{z_{12}^{h_1}\, z_{n-1,n}^{h_n}\, (z_n - q z_1)^{h_1+h_n}}\,\prod_{i=1}^{n-2}\left(\frac{z_{i,i+2}}{z_{i,i+1} z_{i+1,i+2}}\right)^{h_{i+1}},
\ee
and the bare block  is given by 
\begin{multline}
\label{necklacebare}
\cV_{\bf \dl}^{\bf h}({\bm \rho}) = \Big(\prod_{i=1}^n \rho_i^{\dl_i}\Big)\, 
F_N \left[
\begin{array}{l l}
\dl_1+\dl_2-h_1, \dl_2+\dl_3-h_2, ...\, , \dl_n+\dl_1-h_n \\
\qquad \qquad \qquad \quad 2\dl_1, 2\dl_2, ...\, , 2\dl_n
\end{array}\bigg| 
\rho_1, ...\,, \rho_n \right],
\end{multline}
where we introduced a new special function $F_N$ which we call the $n$-point necklace function \eqref{neckalcefunction}. Here, ${\bm \rho} \equiv \{ \rho_1,\rho_2,...\, , \rho_n \}$ are the  torus cross-ratios:
\be
\begin{split}
\label{toruscrossratios}
\rho_1 &= \frac{q \;z_{12}\; z_{n-1,n}}{(z_{n-1}-q z_1)(z_n-q z_2)}  \,, \qquad \rho_2 = \frac{z_{23}(z_n-q z_1)}{z_{13}(z_n-qz_2)} \,, 
\\
\rho_n &= \frac{z_{n-2,n-1} (z_n-q z_1)}{z_{n-2,n} (z_{n-1} -q z_1)}\,,   \qquad  \qquad \rho_i = \frac{z_{i-2,i-1} \; z_{i,i+1}}{z_{i-2,i} \; z_{i-1,i+1}} \quad \mbox{for}\quad 3\leq i \leq n-1 \,.
\end{split}
\ee
Similarly to the 2-point case the torus cross-ratios $\rho_2$, $\rho_n$ arise when plugging  \eqref{cpw4contrnecknfinal} into \eqref{toruscpwblocknfinal} as combinations of $z_i$ and $q$ packed into $\rho_2^{m_2+\dl_2}\rho_n^{m_n+\dl_n}$. This is in contrast to $\rho_1$ and $\rho_i$ for $3\leq i \leq n-1$ which are the plane cross-ratios coming as arguments of the plane conformal blocks. Moreover, $\rho_i$ here  coincide with the plane cross-ratios $\chi_{i-2}$, cf. \eqref{combcross}.  Since the total degree in $z$-coordinates of any ${\rho_i}$ is zero, then one can rewrite them as functions of $u(1)$-invariant ratios $x_i \equiv z_{i+1}/z_i$ (e.g. see \eqref{rho_ratio1} and \eqref{rho_ratio2} in the $n=2$ case). In this way, the bare block is manifestly $u(1)$-invariant. 

Also note that to reproduce the $n=2$ torus block  \eqref{torusblock2} one identifies  $z_3 \equiv  q z_1$ and $z_0 \equiv z_2/q$.  The case of 1-point torus blocks falls out the general formula \eqref{necklaceblock} because the necklace function at $n=1$ can not be reduced to a hypergeometric function ${}_{2}F_1$ with the required  arguments.\footnote{Despite having the relation  \eqref{FNG} there is no such a transformation that brings the $n=1$ necklace function to the 1-point block function.  Nonetheless, we saw that $1$-point  blocks are perfectly well calculated within the torus shadow formalism and, in particular, they can be  obtained from $2$-point torus block by imposing extra  constraints on $\dl_i\,, h_i$, see the end of Section \bref{sec:torus2}.}  Thus, the necklace function defines the torus conformal blocks only when $n=2,3,...\,$. In this respect, the situation is similar to that with the plane conformal blocks where the comb function works properly only when $n=4,5,...\,$. In both cases (plane and torus) one can see that the respective  cross-ratios can be built only having a minimal number of  points not completely fixed by a global conformal symmetry which equals  two in the torus case and four in the plane case.

\section{Conclusion}
\label{sec:concl}

In this paper we have found exact functions of the global torus \cftone $n$-point conformal blocks in the necklace channel. Up to the leg factor these blocks are given by the necklace functions which are hypergeometric-type functions of $n$ torus cross-ratios defined as particular combinations of the locations of the insertion points of primary operators as well as the modular parameter. To this end, we have elaborated the torus shadow formalism which is a version of the standard shadow formalism. Given the (anti-)holomorphic factorization of the \cfttwo torus blocks, the necklace functions also define the \cfttwo global necklace blocks.

Our results can be naturally extended in a few directions. First of all, we note that both the comb functions in plane \cfttwo and the necklace functions in torus \cfttwo belong to the same family of the hypergeometric-type series. It is tempting to speculate that considering  \cfttwo on higher genus-$g$ Riemann surfaces the  conformal block functions in the channels which   generalize the comb and necklace channels are  members of the same hypergeometric-type family of functions. The second natural direction is  to use the elaborated machinery to calculate conformal blocks in thermal CFT$_d$ and their asymptotics, see e.g. \cite{Iliesiu:2018fao,Gobeil:2018fzy,Collier:2019weq,Benjamin:2023qsc}. Third, up to a truncated Virasoro character the global torus blocks are the large-$c$ asymptotics of the Virasoro torus  blocks (the so-called light blocks) \cite{Alkalaev:2016fok,Cho:2017oxl}. Then, the Zamolodchikov's $c$-recursion in torus \cfttwo should be rephrased  in terms of the torus cross-ratios which proved to be an efficient parameterization of torus blocks. Finally, it would be interesting  to find an integral representation for the necklace function along with various Kummer-type and Pfaff-type transformations that could be useful for applications.

\vspace{3mm}

\noindent \textbf{Acknowledgements.}
We are grateful to Daniil Zherikhov, Vladimir Khiteev, and Mikhail Pavlov for  discussions. S.M. was partially supported by the Foundation for the Advancement of Theoretical Physics and Mathematics “BASIS”.

\appendix

\section{Multipoint plane conformal blocks}
\label{app:multiblocks}

The $(n+2)$-point conformal block in the comb channel is found to be \cite{Rosenhaus:2018zqn}
\be
\begin{split}
\label{combblock}
G_{\dl_2,\ldots,\dl_{n}}^{h_{\infty}, h_1,\ldots,h_n, h_0}(z_{\infty},z_1,z_2,...\, ,z_n, z_0) &= \cL^{h_{\infty}, h_1,\ldots,h_n, h_0}(z_{\infty},z_1,z_2,...\, ,z_n, z_0)\\
&\times \, g_{\dl_2,\ldots,\dl_{n}}^{h_{\infty}, h_1,\ldots,h_n, h_0}(\chi_{\infty},\chi_1, ...\,,\chi_{n-3},\chi_0) \,,
\end{split}
\ee
where the $(n+2)$-point leg factor, which has a holomorphic dimension $h_0+\sum_{i=1}^n h_i+h_{\infty}$ and transforms under conformal transformations as an $(n+2)$-point correlation function, is chosen as 
\begin{multline}
\label{combleg}
\cL^{h_{\infty}, h_1,...,h_n, h_0}(z_{\infty},z_1,... ,z_n, z_0) 
= \prod_{i=1}^{n-2}\left(\frac{z_{i,i+2}}{z_{i,i+1}z_{i+1,i+2}}\right)^{h_{i+1}}
\left(\frac{z_{12}}{(z_{\infty}-z_1)(z_{\infty}-z_2)}\right)^{h_{\infty}}\\
\times \, \left(\frac{z_{n-1,n}}{(z_{n-1}-z_0)(z_n-z_0)}\right)^{h_0}
\left(\frac{(z_\infty - z_2)}{(z_{\infty}-z_1)(z_{12})}\right)^{h_1} 
\left(\frac{(z_{n-1}-z_0)}{(z_{n-1,n})(z_n-z_0)}\right)^{h_n},
\end{multline}
the $(n+2)$-point bare conformal block in the comb channel is given by
\begin{multline}
\label{combbare}
g_{\dl_2,\ldots,\dl_{n}}^{h_{\infty},h_1,...,h_n,h_0}(\chi_{\infty},\chi_1,...,\chi_{n-3},\chi_0) = \chi_{\infty}^{\dl_2} \chi_1^{\dl_3} \ldots \chi_{n-3}^{\dl_{n-1}} \chi_0^{\dl_n} \\ 
F_K \left[ 
\begin{array}{l l}
h_\infty+\dl_2-h_1,\dl_2+\dl_3-h_2,..., \dl_{n-1}+\dl_n -h_{n-1}, \dl_n + h_0 - h_n\\
\qquad \qquad \qquad \qquad \qquad 2\dl_2,..., 2\dl_n
\end{array}\bigg| \chi_{\infty}, \chi_{1},..., \chi_{n-3},\chi_{0}
\right],
\end{multline}
where the cross-ratios are 
\be
\label{combcross}
\chi_{\infty} = \frac{(z_{\infty}-z_1) z_{23}}{(z_{\infty}-z_2) z_{13}} \,, \quad  \chi_{0} = \frac{z_{n-2,n-1} (z_n-z_0)}{z_{n-2,n} (z_{n-1}-z_0)} \,, \quad \chi_i = \frac{z_{i,i+1} z_{i+2,i+3} }{z_{i,i+2} z_{i+1,i+3} } \,, \quad 1\leq i \leq n-3 \,,
\ee
and $F_K$ is the  comb function \cite{Rosenhaus:2018zqn}
\begin{multline}
\label{combfunction}
F_K \left[
\begin{array}{l l}
a_1,b_1,...\,,b_{n-1},a_2\\
\qquad c_1,...\,, c_n
\end{array} \bigg| x_1,...\,, x_n \right]  
\\ = 
\sum_{k_1,\ldots,k_n = 0}^{\infty} \frac{(a_1)_{k_1} (b_1)_{k_1+k_2} (b_2)_{k_2+k_3}...(b_{n-1})_{k_{n-1}+k_n} (a_2)_{k_n}}{(c_1)_{k_1}... (c_n)_{k_n}} \frac{x_1^{k_1}}{k_1!}\,...\,\frac{x_n^{k_n}}{k_n!} \,.
\end{multline}

\section{Necklace function $F_N$}
\label{app:necklacefunction}

\subsection{Definition}

The necklace function is defined to be 
\be
\label{neckalcefunction}
F_N \left[
\begin{array}{l l}
a_1, ..., a_n \\
c_1, ..., c_n
\end{array}\bigg| 
x_1,..., x_n 
\right]
= \sum_{m_1,... ,m_n = 0}^{\infty} \frac{(a_1)_{m_1+m_2} (a_2)_{m_2+m_3} ... (a_n)_{m_n+m_1} }{(c_1)_{m_1} ... (c_n)_{m_n} } \frac{x_1^{m_1} }{m_1!} \,...\, \frac{x_n^{m_n} }{m_n!} \,,
\ee
where $n=1,2,...\,$. At $n=1$ it is given by 
\be
\label{FNG}
F_N \left[
\begin{array}{l l}
a_1  \\
 c_1
\end{array}\bigg| 
x_1
\right] 
= 
{}_2 F_1 
\left[
\begin{array}{l l}
\dps\frac{a_1}{2} , \frac{1+a_1}{2} \\
\hspace{6mm}  c_1
\end{array}\bigg| \; 4x_1
\right],
\ee
which is the hypergeometric function.  At $n=2$ it is given by  
\be
\label{appelfunction4}
F_N \left[
\begin{array}{l l}
a_1, a_2  \\
 c_1, c_2
\end{array}\bigg| 
x_1, x_2
\right] 
= 
\sum_{m_1, m_2 = 0}^{\infty} \frac{(a_1)_{m_1+m_2} (a_2)_{m_2+m_1}  }{(c_1)_{m_1} (c_2)_{m_2} } \frac{x_1^{m_1} }{m_1!} \frac{x_2^{m_2} }{m_2!} \,,
\ee
which is the fourth Appell function $F_4$. 

\subsection{Differential equations}
\label{app:B2}

We write the necklace function \eqref{neckalcefunction} as
\be
\label{necklacecoeffs}
F_N \left[
\begin{array}{l l}
a_1, ...\,, a_n \\
c_1, ...\,, c_n
\end{array}\bigg| 
x_1,...\,, x_n 
\right] = \sum_{m_1,\ldots ,m_n = 0}^{\infty} A_{m_1,\ldots, m_n} x_1^{m_1}...\, x_n^{m_n} \,.
\ee
From the definition of the series coefficients $A_{m_1,\ldots, m_n}$ one finds that they satisfy the following  recursion relations:
\be
\label{necklacerecursion1}
A_{m_1+1,m_2,\ldots ,m_n} = \frac{(a_1+m_1+m_2) (a_n+m_n+m_1)}{(c_1+m_1)(m_1+1)} A_{m_1,m_2,\ldots ,m_n} \,,
\ee
\be
\label{necklacerecursionk}
A_{m_1,\ldots ,m_k+1, \ldots ,m_n} = \frac{(a_{k-1}+m_{k-1}+m_k) (a_k+m_k+m_{k+1})}{(c_k+m_k)(m_k+1)} A_{m_1,\ldots ,m_k, \ldots ,m_n} \,, \quad 1<k<n \,,
\ee
\be
\label{necklacerecursionlast}
A_{m_1,m_2,\ldots ,m_n+1} = \frac{(a_{n-1}+m_{n-1}+m_n) (a_n+m_n+m_1)}{(c_n+m_n)(m_n+1)} A_{m_1,m_2,\ldots ,m_n} \,.
\ee
Introducing  the counting operators
\be
\label{countingops}
N_i = x_i \frac{\partial}{\partial x_i} \,, \qquad i=1,...\, ,n \,,
\ee
with $x_i$ being arguments of $F_N$ in \eqref{necklacecoeffs} one can show that the system of $n$ recursion relations \eqref{necklacerecursion1}-\eqref{necklacerecursionlast} can be equivalently represented as $n$ PDEs 
\be
\label{necklaceequation1}
\left[x_1 (N_1+N_2+a_1)(N_1+N_n+a_n) - N_1(N_1+c_1-1)\right] F_N = 0 \,,
\ee
\be
\label{necklaceequationk}
\left[x_k(N_{k-1}+N_k+a_{k-1})(N_k+N_{k+1}+a_k)-N_k(N_k+c_k-1)\right]F_N = 0 \,, \quad 1<k<n \,,
\ee
\be
\label{necklaceequationlast}
\left[x_n(N_{n-1}+N_n+a_{n-1})(N_n+N_1+a_n)-N_n(N_n+c_n-1)\right] F_N = 0 \,.
\ee
At $n=2$ the equation system   \eqref{necklaceequation1}-\eqref{necklaceequationlast} can be cast into the form
\be
\label{appellequations4}
\ba{c}
\dps
\left[ x_1(1-x_1) \partial_1^2 -x_2^2 \partial_2^2 - 2 x_1 x_2 \partial_1 \partial_2 +(c_1-(a_1+a_2+1)x_1)\partial_1 -(a_1+a_2+1)\partial_2 - a_1 a_2 \right] F_N=0 \,, 
\vspace{2mm}
\\
\dps
\left[ x_2(1-x_2) \partial_2^2 -x_1^2 \partial_1^2 - 2 x_1 x_2 \partial_1 \partial_2 +(c_2-(a_1+a_2+1)x_2)\partial_2 -(a_1+a_2+1)\partial_1 - a_1 a_2 \right] F_N=0 \,,
\ea
\ee
which is a system of PDEs on the fourth Appell function $F_4$ \eqref{appelfunction4} \cite{Bateman:100233}. Its general solution is given by \cite{Bateman:100233}
\be
\ba{l}
\label{generalsolappell}
\dps
F_4(x_1,x_2)=C_1\,F_4 \left[
\begin{array}{l l}
a_1, a_2  \\
c_1, c_2
\end{array}\bigg| 
x_1, x_2
\right] 
\vspace{2mm}
\\
\dps
\hspace{19mm}+C_2\,x_1^{1-c_1}\,F_4\left[
\begin{array}{l l}
a_1-c_1+1, a_2-c_1+1  \\
\qquad \quad 2-c_1, c_2
\end{array}\bigg| 
x_1, x_2
\right] 
\vspace{2mm}
\\
\dps
\hspace{19mm}+C_3\,x_2^{1-c_2}\,F_4\left[
\begin{array}{l l}
a_1-c_2+1, a_2-c_2+1  \\
\qquad \quad c_1, 2-c_2
\end{array}\bigg| 
x_1, x_2
\right] 
\vspace{2mm}
\\
\dps
\hspace{19mm}+C_4\,x_1^{1-c_1}\,x_2^{1-c_2}\,F_4\left[
\begin{array}{l l}
2+a_1-c_1-c_2, 2+a_2-c_1-c_2  \\
\qquad \qquad 2-c_1, 2-c_2
\end{array}\bigg| 
x_1, x_2
\right],
\ea
\ee
where $C_i$, $i=1,2,3,4$ are integration constants. The first line defines the necklace block \eqref{torusbare2}, while the others define the shadow block (see below).

\paragraph{Connection with the torus Casimir equations.} When considering the torus conformal blocks the system \eqref{necklaceequation1}-\eqref{necklaceequationlast} has to be regarded as the torus Casimir equations. For example, the Casimir equations which  define the 2-point necklace block reads as \cite{Alkalaev:2022kal} 
\be
\label{casimireqs2}
\ba{l}
\dps
\left[-q^2\partial_q^2 + \frac{2q}{1-q}q\partial_q - \frac{q}{(1-q)^2}\left(\cL_{-1}^{(1)}+\cL_{-1}^{(2)}\right)\left(\cL_{1}^{(1)} + \cL_{1}^{(2)}\right) + \dl_1 (\dl_1-1) \right] \cF_{\dl_1,\dl_2}^{h_1,h_2}(q,z_1,z_2) = 0 \,, 
\vspace{2mm}
\\
\dps
\left[- q^2\partial_q^2+\frac{2q}{1-q}q\partial_q - \frac{1}{(1-q)^2}\left(\cL_{-1}^{(1)}+q\cL_{-1}^{(2)}\right)\left(\cL_1^{(2)}+q\cL_1^{(1)}\right) +\frac{1+q}{1-q}\cL_0^{(2)}
\right.
\vspace{2mm}
\\
\dps
\hspace{50mm}\left. -\left(\cL_0^{(2)}\right)^2 - 2\cL_0^{(2)}q\partial_q + \dl_2 (\dl_2 -1) \right] \cF_{\dl_1,\dl_2}^{h_1,h_2}(q,z_1,z_2)
= 0 \,,
\ea
\ee
where $\cL^{(i)}_m$ are given in  \eqref{cL}.  Substituting here the 2-point necklace block \eqref{torusblock2} and stripping off the leg factor \eqref{torusleg2} one finds that this system transforms into the Appell  system \eqref{appellequations4} provided that $c_i\rightarrow 2 \dl_i$, $a_i \rightarrow \dl_1+\dl_2-h_i$, and $x_i \rightarrow \rho_i$, where the torus cross-ratios $\rho_i$ are given by \eqref{rho_ratio1} and \eqref{rho_ratio2}. Thus, the general solution  to the Appell equations \eqref{appellequations4} and the Casimir equations \eqref{casimireqs2} is the same and is given by \eqref{generalsolappell}: in the first line one recognizes the necklace block, while the other three lines contain the shadow blocks  obtained from the necklace block by $\dl_i \rightarrow 1-\dl_i$ ($a_{i}, c_{i}$ will be respectively changed).

It would be important to verify that the Casimir system \cite{Alkalaev:2022kal} does transform into the necklace system \eqref{necklaceequation1}-\eqref{necklaceequationlast} for any $n$. We have checked this statement up to $n=4$.

\providecommand{\href}[2]{#2}\begingroup\raggedright\endgroup

\end{document}